\documentclass[11pt]{article}
\usepackage{amsfonts,amsmath,amssymb}
\usepackage{enumerate}
\usepackage{hyperref}
\usepackage{bbm}
\usepackage{nicefrac}
\usepackage[all]{xy}
\usepackage{graphicx}
\usepackage{bm}
\usepackage{makecell}
\usepackage[table]{xcolor}

\usepackage[utf8]{inputenc}

\usepackage{upgreek}
\usepackage{cancel}			

\usepackage{framed}

\usepackage{longtable}      

\usepackage{booktabs}
\newcommand{\ra}[1]{\renewcommand{\arraystretch}{#1}}

\newcommand{\be}{\begin{equation}}
\newcommand{\ee}{\end{equation}}

\newcommand{\bea}{\begin{eqnarray}}
\newcommand{\eea}{\end{eqnarray}}

\newcommand{\bes}{\begin{subequations}}
\newcommand{\ees}{\end{subequations}}

\newcommand{\cN}{{\cal N}}

\def\sst#1{{\scriptscriptstyle #1}}

\def\0{{\sst{(0)}}}
\def\1{{\sst{(1)}}}
\def\2{{\sst{(2)}}}
\def\3{{\sst{(3)}}}
\def\4{{\sst{(4)}}}
\def\5{{\sst{(5)}}}
\def\6{{\sst{(6)}}}
\def\7{{\sst{(7)}}}
\def\8{{\sst{(8)}}}

\def\m{{\sst{(m)}}}
\def\n{{\sst{(n)}}}
\def\k{{\sst{(k)}}}
\def\neqone{{\sst{(n=1)}}}
\def\keqone{{\sst{(k=1)}}}

\def\cM{{{\cal M}}}

\allowdisplaybreaks  

\usepackage{multirow}
\usepackage{rotating}

\newcommand{\ba}{\begin{align}}
\newcommand{\ea}{\end{align}}

\newcommand{\bse}{\begin{subequations}}
\newcommand{\ese}{\end{subequations}}

\allowdisplaybreaks

\global\long\def\cn{\mathcal{N}}

\global\long\def\cy{\mathcal{Y}}

\global\long\def\sofour{\text{SO}(4)}

\begin{document}

\makeatletter
\renewcommand{\theequation}{\thesection.\arabic{equation}}
\@addtoreset{equation}{section}
\makeatother

\begin{titlepage}

\begin{flushright}
IFT-UAM/CSIC-19-157 \\
%
\end{flushright}

\vspace{5pt}

   \begin{center}
   \baselineskip=16pt

   \begin{Large}\textbf{
\mbox{Universal properties of Kaluza-Klein gravitons}
}
   \end{Large}

\vspace{25pt}

{\large  Kevin Dimmitt$^{1}$ ,\ Gabriel Larios$^{2}$ ,\, Praxitelis Ntokos$^{1}$ \,and \,  Oscar Varela$^{1,2}$}
		
\vspace{25pt}

	\begin{small}

	{\it $^{1}$ Department of Physics, Utah State University, Logan, UT 84322, USA}  \\

	\vspace{15pt}
	
	  {\it $^{2}$ Departamento de F\'\i sica Te\'orica and Instituto de F\'\i sica Te\'orica UAM/CSIC , \\
   Universidad Aut\'onoma de Madrid, Cantoblanco, 28049 Madrid, Spain}   \\	
		
	\end{small}

\vskip 50pt

\end{center}

\begin{center}
\textbf{Abstract}
\end{center}

\begin{quote}

Vacua of different gaugings of $D=4$ ${\cal N}=8$ supergravity that preserve the same supersymmetries and bosonic symmetry tend to exhibit the same universal mass spectrum within their respective supergravities. For AdS$_4$ vacua in gauged supergravities that arise upon consistent truncation of string/M-theory, we show that this universality is lost at higher Kaluza-Klein levels. However, universality is still maintained in a milder form, at least in the graviton sector: certain sums over a finite number of states remain universal. Further, we derive a mass matrix for Kaluza-Klein gravitons which is valid for all the AdS$_4$ vacua in string/M-theory that uplift from the gaugings of $D=4$ ${\cal N}=8$ supergravity that we consider. The mild universality of graviton mass sums is related to the trace of this mass matrix.

\end{quote}

\vfill

\end{titlepage}

\tableofcontents



\section{Introduction}


It is now well established that maximal gauged supergravities in four dimensions typically come in one-parameter families \cite{Dall'Agata:2012bb,Dall'Agata:2014ita,Inverso:2015viq}. All members in a given family share the same gauge group. The parameter, discrete or continuous, that characterises the family leaks in as a remnant of the freedom of the ungauged theory to select an electric/magnetic duality frame before the gauging is introduced. It also happens that different such gaugings usually have vacua, of anti-de Sitter (AdS), de Sitter or Minkowski type, that exhibit the same residual supersymmetries $0 \leq \cN \leq 8$ and the same residual bosonic symmetry \cite{Dall'Agata:2012bb,DallAgata:2011aa,Borghese:2012qm,Borghese:2012zs,Borghese:2013dja,Gallerati:2014xra}. Often (though not always), vacua with the same (super)symmetries have identical mass spectra within their correspoding gauged supergravities. An interesting question therefore arises for $D=4$ $\cN=8$ gaugings with a higher dimensional origin: for vacua of different such $\cN=8$ gaugings with the same symmetry and the same spectrum within their corresponding $\cN=8$ theories, are the masses still the same up the corresponding Kaluza-Klein (KK) towers?

In this paper, we will address this question by focusing on three $\cN=8$ gaugings with AdS vacua: the purely electric SO(8) gauging \cite{deWit:1982ig}, the dyonic ISO(7) gauging \cite{Guarino:2015qaa} and the dyonic $\big(\textrm{SO}(6)\times\textrm{SO}(1,1)\big)\ltimes\mathbb{R}^{12}$  gauging as described in \cite{DallAgata:2011aa}. All three gaugings enjoy higher dimensional origins in, respectively, $D=11$ supergravity \cite{Cremmer:1978km} on $S^7$ \cite{deWit:1986iy}, massive type IIA supergravity \cite{Romans:1985tz} on $S^6$ \cite{Guarino:2015jca,Guarino:2015vca} and type IIB supergravity \cite{Schwarz:1983qr} on an S-fold geometry \cite{Inverso:2016eet}. In order to narrow down and simplify the problem, we will further focus on the SU(3)--invariant sector of these gaugings. This sector has been explicitly uplifted to the respective higher-dimensional supergravities \cite{Larios:2019kbw,Varela:2015uca,Guarino:2019oct}, and the corresponding vacua have been charted \cite{Warner:1983vz,Guarino:2015qaa,Guarino:2019oct}. The list of possible symmetries of AdS vacua in this sector across all three gaugings of interest has been summarised for convenience in table \ref{tab:SolutionsinGaugings}. Only the symmetries are indicated in the table, regardless of their actual embedding into the relevant gauge groups and ultimately E$_{7(7)}$. For example, the same table row accounts for the $\textrm{SU}(3) \times \textrm{U}(1)_c$ and $\textrm{SU}(3) \times \textrm{U}(1)_v$ solutions of SO(8) and ISO(7) supergravity, which differ in their  embeddings into E$_{7(7)}$. The table also shows whether vacua with the same (super)symmetries in different theories exhibit the same mass spectrum within their corresponding $D=4$ $\cN=8$ theories. All of them do, except for the solution with $\cN=0$ SU(3) symmetry. For this reason, we will not be concerned with the latter in this paper.
 
\begin{table}[]
\centering
\resizebox{\textwidth}{!}{
\begin{tabular}{lcccc}
\hline
(Super)symmetry               
& $\qquad\textrm{SO}(8) \qquad$                
& $\qquad\textrm{ISO}(7) \qquad$ 
&$\big(\textrm{SO}(6)\times\textrm{SO}(1,1)\big)\ltimes\mathbb{R}^{12}$ 
& same spectrum?
\\ \hline \hline
$\cN=8 \ , \; \textrm{SO}(8)  $
&$\checkmark $
& $\times$
& $\times$
& --
\\ \hline
$\cN=2 \ , \; \textrm{SU}(3) \times \textrm{U} (1)  $
&$\checkmark $
& $\checkmark $
& $\times$
& $\checkmark $
\\ \hline
$\cN=1 \ , \; \textrm{G}_2  $
&$\checkmark $
& $\checkmark $
& $\times$
& $\checkmark $
\\ \hline
$\cN=1 \ , \;  \textrm{SU}(3) $
&$\times$
& $\checkmark $
& $\checkmark $
& $\checkmark $
\\ \hline
$\cN=0 \ , \;  \textrm{SO}(7) $
& $\checkmark $
& $\checkmark $
& $\times $
& $\checkmark $
\\ \hline
$\cN=0 \ , \;  \textrm{SO}(6) $
& $\checkmark $
& $\checkmark $
& $\checkmark $
& $\checkmark $
\\ \hline
$\cN=0 \ , \;  \textrm{G}_2 $
& $\times  $
& $\checkmark $
& $\times  $
&  --  
\\ \hline
$\cN=0 \ , \;  \textrm{SU}(3) $
& $\times  $
& $\checkmark $
& $\checkmark $
& $\times  $ 
\\ \hline
\end{tabular}
}

\caption{\footnotesize{Possible residual (super)symmetries, regardless of their E$_{7(7)}$ embedding, of AdS vacua  in the $\textrm{SU}(3)$-invariant sector of the three different gaugings that we consider.}\normalsize}
\label{tab:SolutionsinGaugings}
\end{table}

It was already shown in \cite{Pang:2017omp} that the $\cN=2$ $\textrm{SU}(3) \times \textrm{U}(1)$-invariant solutions of the SO(8) \cite{Warner:1983vz} and ISO(7) gaugings \cite{Guarino:2015jca} do cease to have the same KK spectrum upon uplift to $D=11$ \cite{Corrado:2001nv} and type IIA \cite{Guarino:2015jca}, thus answering in the negative the question posed above. Fortunately, it was not necessary to compute the entire KK spectrum to elucidate that question in \cite{Pang:2017omp}. Computing the spectrum of gravitons following \cite{Bachas:2011xa} was enough to give an answer. In this paper, we will extend this statement to all solutions with at least SU(3) symmetry. Any pair of such solutions in table \ref{tab:SolutionsinGaugings} with the same (super)symmetry and the same spectrum within their $\cN=8$ supergravities fails to have the same spectrum of KK gravitons. This is so regardless of the embedding of the residual symmetry group into the gauge group, and ultimately into the duality group E$_{7(7)}$ of the ungauged $\cN=8$ theory. For example, both the SO$(7)_v$ and SO$(7)_c$ solutions of the SO(8) gauging have the same spectrum within the $\cN=8$ theory \cite{Bobev:2010ib}. But, as we will show in this paper, the spectrum of gravitons for both solutions differ.

It was also observed in \cite{Pang:2017omp} that, despite their different KK mode structure, certain sums of masses up the Kaluza-Klein tower did still remain universal for the $\cN=2$ $\textrm{SU}(3) \times \textrm{U}(1)$ solutions of the SO(8) and ISO(7) gaugings. In other words, while the eigenvalue-by-eigenvalue equivalence of both spectra at the lowest KK level was lost up the KK tower, certain combinations of higher KK modes did still remain universal. In section \ref{sec: SpecMth}, we will find the same behaviour for solutions of the SO(8) and ISO(7) gauging within the SU(3)-invariant sector. The solutions of the $\big(\textrm{SO}(6)\times\textrm{SO}(1,1)\big)\ltimes\mathbb{R}^{12}$ gauging behave similarly but slightly differently. For that reason, we relegate their discussion to the concluding section \ref{sec:Discussion}. In \cite{Pang:2017omp}, the relevant combinations of KK modes were identified as traces of a KK graviton mass matrix. In section \ref{sec:KKGravMassMat}, we formalise this notion and introduce an SL(8)-covariant KK graviton mass matrix whose form is qualitatively similar to the bosonic mass matrices of $D=4$ $\cN=8$ gauged supergravity (see \cite{Trigiante:2016mnt}). Prior to discussing this, we complete in section \ref{sec:IIBSO4Spectrum} the KK graviton spectrum for all known solutions of the $\big(\textrm{SO}(6)\times\textrm{SO}(1,1)\big)\ltimes\mathbb{R}^{12}$ gauging by computing the spin-2 spectrum of a solution with $\cN=4$ supersymmetry and SO(4) symmetry \cite{Inverso:2016eet}, which lies outside the SU(3) sector of section \ref{sec: SpecMth}. 


In this paper, we want to compute the spectrum of massive gravitons about the AdS$_4$ solutions of the ten- and eleven-dimensional supergravities specified above. For later reference, the relevant geometries are of the form
\begin{equation}	\label{eq:Mthbkg}
	d\hat{s}_{d+4}^2=e^{2A(y)}\Big[(\bar{g}_{\mu\nu}(x)+h_{\mu\nu}(x,y))dx^\mu dx^\nu+d\bar{s}_d^2(y)\Big]\;,
\end{equation}
where $d=7$ in M-theory and $d=6$ in type II. The Einstein frame is used in the latter case. The external and internal coordinates have been collectively denoted $(x,y)$. The metrics $\bar{g}_{\mu\nu}\,dx^\mu dx^\nu \equiv ds^2 (\textrm{AdS}_4)$ and $d\bar{s}_d^2(y)$ respectively denote the unit-radius four-dimensional anti-de Sitter metric, and a background metric on the internal $d$-dimensional space that will be specified below on a case-by-case basis. The warp factor $ e^{2A(y)}$ takes values on the internal space. Finally, $h_{\mu\nu}$ is taken to be a spin-2 perturbation over $\textrm{AdS}_4$ that also depends on the internal coordinates. More concretely, the perturbation is assumed to take on the factorised form 
\begin{equation}
	h_{\mu\nu}(x,y) =h^{[tt]}_{\mu\nu}(x)\cy(y) \; , 
\end{equation}
with $\cy(y)$ a function on the internal space only, and $h^{[tt]}$ transverse ($\bar{\nabla}^\mu h^{[tt]}_{\mu\nu}$= 0) with respect to the Levi-Civita connection corresponding to $\bar{g}_{\mu\nu}$, traceless ($\bar{g}^{\mu\nu}h^{[tt]}_{\mu\nu}=0$), and subject to the Fierz-Pauli equation
\begin{equation}
	\bar{\square}h^{[tt]}_{\mu\nu}=(M^2L^2-2)h^{[tt]}_{\mu\nu} \; ,
\end{equation}
for a graviton of squared mass $M^2$. Here, $L$ is the effective AdS$_4$ radius (introduced in our context by the warping $e^{2A(y)}$), such that the combination $M^2L^2$ is dimensionless. 

While the computation of supergravity spectra in general is a very complicated problem, the calculation of massive graviton spectra is comparatively much simpler. The reason is that the linearised spin-2 equations decouple from the supergravity fluxes and become a boundary value problem involving only the warp factor and the internal background metric. Indeed, with the above assumptions, the linearised ten- and eleven-dimensional Einstein equations reduce to the eigenvalue problem \cite{Bachas:2011xa}
\begin{equation}	\label{eq: GeneralPDE}
	-\frac{e^{-(d+2)A}}{\sqrt{\bar{g}}}\partial_M\big(e^{(d+2)A}\sqrt{\bar{g}}\,\bar{g}^{MN}\partial_N \cy \big)=M^2L^2\cy\,,
\end{equation}
with $\bar{g}^{MN}$, $M,N=1,\dots,d\,,$ the inverse metric and $\bar{g}$ the determinant of the internal metric $d\bar{s}_d^2$ in \eqref{eq:Mthbkg}. For unwarped geometries, $A=0$, (\ref{eq: GeneralPDE}) reduces to the spectral problem for the Laplacian on the internal space. Previous calculations of KK graviton spectra in related contexts include \cite{Klebanov:2009kp,Bachas:2011xa,Richard:2014qsa,Passias:2016fkm,Passias:2018swc,Gutperle:2018wuk,Chen:2019ydk,Speziali:2019uzn,Andriot:2019hay}.


\section{Massive gravitons with at least SU(3) symmetry} \label{sec: SpecMth}

\addtocontents{toc}{\setcounter{tocdepth}{2}}


We now compute the KK graviton spectrum about the AdS$_4$ solutions of $D=11$ supergravity, massive type IIA supergravity and type IIB supergravity that uplift from critical points of SO(8) supergravity, dyonic ISO(7) supergravity and $\big(\textrm{SO}(6)\times\textrm{SO}(1,1)\big)\ltimes\mathbb{R}^{12}$ supergravity with at least SU(3) symmetry. For convenience, some features of the SU(3)-invariant sector of these $D=4$ $\cN=8$ gaugings are summarised in appendix \ref{sec:SU3sector}.


\subsection{M-theory} \label{sec:Mtheory}

The class of AdS$_4$ solutions of $D=11$ supergravity \cite{Cremmer:1978km} that we are interested in arises upon consistent uplift on $S^7$ \cite{deWit:1986iy} of critical points of $D=4$ $\cN=8$ SO(8)-gauged supergravity \cite{deWit:1982ig}. For definiteness, we will restrict to critical points that preserve at least the SU(3) subgroup of SO(8). There are six such vacua \cite{Warner:1983vz}. The  corresponding uplifts are given by the $D=11$ solutions first found in \cite{Freund:1980xh,Corrado:2001nv,deWit:1984nz,deWit:1984va,Englert:1982vs,Pope:1984bd}. These solutions are invariant, both in $D=4$ and in $D=11$, under a number of subgroups of SO(8) larger than SU(3), and display supersymmetries $\cN=0,1,2,8$. See table \ref{tab:SolutionsinGaugings} for a summary. The entire spectrum about the Freund-Rubin $\cN=8$ SO(8)--invariant AdS$_4$ solution \cite{Freund:1980xh} has long been known \cite{Englert:1983rn,Sezgin:1983ik,Biran:1983iy}  (see also \cite{Duff:1986hr} for a review). The spectrum of gravitons about the $\cN=2$ $\textrm{SU}(3) \times \textrm{U}(1)_c$--invariant solution \cite{Corrado:2001nv} is also known \cite{Klebanov:2009kp}. The graviton spectra that we will give below for the four other AdS$_4$ solutions in this sector are new.

A convenient starting point for our analysis is the local geometries recently presented in \cite{Larios:2019kbw}. In that reference, the full, dynamical SU(3)--invariant sector of $D=4$ $\cN=8$ SO(8)-gauged supergravity \cite{deWit:1982ig} was uplifted to $D=11$ using the consistent truncation formulae of \cite{Varela:2015ywx}. In particular, the results of \cite{Larios:2019kbw} provide a unified treatment for all the $D=11$ AdS$_4$ solutions that uplift from critical points of $D=4$ $\cN=8$ SO(8)-gauged supergravity with at least SU(3) symmetry. In order to simplify the calculations, we will focus on two disjoint further subsectors with symmetries G$_2$ and SU$(4)_c$ larger than SU(3). We will obtain the graviton spectra for arbitrary constant values of the $D=4$ scalars in those sectors. Finding the actual spectra about each individual solution will simply entail an evaluation of those formulae at the corresponding scalar vevs.

%
%

\subsubsection{Massive gravitons with at least G$_2$ symmetry} \label{sec:G2sec}

The G$_2$-invariant sector of the $D=4$ SO(8) supergravity contains an $\textrm{SL}(2, \mathbb{R})/\textrm{SO}(2)$ dilaton-axion pair $(\varphi , \chi)$, in the notation of appendix \ref{sec:SU3sector}. The $D=11$ uplift of this sector was given in section 3.2.3 of \cite{Larios:2019kbw}. The warp factor and internal $d=7$ geometry that feature in (\ref{eq:Mthbkg}) are given by 
\begin{equation} \label{eq:metG2}
    e^{2A}=e^{-\varphi}X^{1/3}\Delta_{1}^{2/3}L^2 \; , \qquad     d\Bar{s}_7^2=g^{-2}L^{-2} \Big( e^{3\varphi}X^{-3}d\beta^2+e^{\varphi}\Delta_1^{-1}\sin^2{\beta}\,ds^2(S^6) \Big) \, .
\end{equation}
Here $\beta$ is an angle on $S^7$, with\footnote{This range of $\beta$ corrects a typo below (B.22) of \cite{Larios:2019kbw}.} $0\leq \beta \leq \pi$, and $ds^2(S^6)$ is the round Einstein metric on the unit radius $S^6$. The dilaton $\varphi$ appears explicitly in (\ref{eq:metG2}) and the axion $\chi$ appears through the combinations $X$ defined in equation (\ref{scalDefs}) and
\begin{equation}
\Delta_1  = X \big(e^{2\varphi}\sin^2\beta+e^{-2\varphi}X^2 \cos^2\beta \big) \, .
\end{equation}
Finally, $g$ and $L$ are constants. The former is the gauge coupling of the $D=4$ supergravity and the latter is related via (\ref{eq:AdSLV}) to the G$_2$--invariant potential $V$, given by (\ref{eq:scalarpotSU3inSO8}) with the identifications (\ref{eq:G2sector}). The geometry (\ref{eq:metG2}) is in fact invariant  under the $\textrm{SO}(7)_v$ that rotates the round $S^6$. When $\chi \neq 0$, the symmetry of the full $D=11$ configuration is broken to G$_2$ by the supergravity four-form field strength.

For the class of geometries (\ref{eq:metG2}), the differential equation (\ref{eq: GeneralPDE}) becomes
\begin{equation} \label{eq:G2PDE}
\Big[ e^{-3\varphi}X^3(\partial_\beta^2+6\cot{\beta}\partial_\beta)+e^{-\varphi}\Delta_1 \sin^{-2}{\beta}\,\square_{S^6} \Big] \cy = -g^{-2} M^2\, \cy \, ,
\end{equation}
where  $\square_{S^6}$ is the $S^6$ Laplacian. Using separation of variables,
\begin{equation} \label{EFansatz1}
\cy=f \, \cy_k \; , 
\end{equation}
where $f= f(\beta)$ depends only on $\beta$ and $\cy_k$ are the $S^6$ spherical harmonics, 
\begin{equation}
\square_{S^6}\cy_k = -k(k+5)\cy_k \; , 
\end{equation}
the PDE (\ref{eq:G2PDE}) reduces to an ODE for $f(\beta)$,
\begin{equation} \label{eq:G2ODE}
    e^{-3\varphi}X^3(f''(\beta)+6\cot{\beta}f'(\beta))-e^{-\varphi}\Delta_1\sin^{-2}{\beta}\,k(k+5)f(\beta)=-g^{-2}M^2f(\beta) \, ,
\end{equation}
where a prime denotes derivative with respect to $\beta$. Finally, it is convenient to introduce a further change of variables,
\begin{equation} \label{Changvars}
u=\cos^2\beta \; , \qquad 
f(u)=(1-u)^{\frac{k}{2}}  H(u) \; .
\end{equation}
The independent variable $u$ now ranges in $0 \leq u \leq 1$, covering this interval twice given the range of $\beta$ below (\ref{eq:metG2}). In the variables (\ref{Changvars}), the differential equation (\ref{eq:G2ODE}) takes on the standard hypergeometric form
\begin{equation} \label{eq:HGeq}
u(1-u)H''+ (c-(1+a_+ +a_- )u)H' - a_+ a_- H = 0 \; , 
\end{equation}
with
\begin{equation} \label{eq:HGparams}
    a_\pm = \tfrac{1}{2}(k+3) \mp \tfrac{1}{2}e^{3\varphi/2}X^{-3/2}\sqrt{  g^{-2}M^2+9e^{-3\varphi}X^3+k(k+5)(e^{-3\varphi}X^3-e^{\varphi}X) } \; , 
    \hspace{0.5cm}
    c = \tfrac{1}{2}\, .
\end{equation}

The two linearly independent solutions to (\ref{eq:HGeq}) are given by the hypergeometric functions
\be \label{LinIndepeHGSols}
 {}_2F_1(a_+,a_- ,c ; u)\, \quad {\rm and}\quad  u^{1-c}{}_{\, 2} F_1(1+a_+-c,1+a_--c,2-c ; u) \; .
\ee
Both solutions are regular at $u = 0$ for all values of the parameters (\ref{eq:HGparams}). At $u=1$, however, regularity imposes restrictions on the parameters. Regularity of the first solution in (\ref{LinIndepeHGSols}) demands $a_+ = - j $ with $j$ a non-negative integer. Bringing this condition to (\ref{eq:HGparams}), we find a first tower of KK graviton squared masses: 
\begin{equation} \label{KKevenbranch}
 g^{-2}M_{\1 \, j,k}^2=e^{-3\varphi}X^3(2j+k)(2j+k+6)+e^{-\varphi}X(e^{2\varphi}-e^{-2\varphi}X^2)k(k+5) \, .
\end{equation}
The corresponding eigenfunctions are given by (\ref{EFansatz1}), (\ref{Changvars}), with $H(u)$ given by the first choice in (\ref{LinIndepeHGSols}), namely
\begin{equation} \label{EigenFunceven}
    \cy_{\1 \, j,k} = \cy_k \, \sin^{k}\beta \sum_{s=0}^{j}(-1)^s \binom{j}{s}\frac{(j+k+3)_s}{(\frac12)_s}\cos^{2s}\beta 
\end{equation}
(no sum in $k$), where
\begin{equation} \label{Pochhammer}
(x)_s =
\left\{
\begin{array}{lll}
1 & ,                        & \textrm{if $s = 0$}  \\
x (x+1) \cdots (x+s-1)  & ,
& \textrm{if  $s  > 0$ }
 \end{array} \right.
\end{equation}
is the Pochhammer symbol. Regularity of the second solution in (\ref{LinIndepeHGSols}) at $u=1$ in turn requires $1+a_+ -c = - j $, with $j$ again a non-negative integer. Bringing this condition to (\ref{eq:HGparams}), we find a second tower of KK graviton squared masses:
\begin{equation} \label{KKoddbranch}
 g^{-2}M_{\2 \, j,k} ^2=e^{-3\varphi}X^3(2j+1+k)(2j+1+k+6)+e^{-\varphi}X(e^{2\varphi}-e^{-2\varphi}X^2)k(k+5) \; .
\end{equation} 
The associated eigenfunctions are now given by (\ref{EFansatz1}), (\ref{Changvars}), with $H(u)$ given by the second choice in (\ref{LinIndepeHGSols}):
\begin{eqnarray} \label{EigenFuncodd}
  \cy_{\2 \, j,k} = \cy_k \, \sin^{k}\beta \sum_{s=0}^{j}(-1)^s \binom{j}{s}\frac{(j+k+4)_s}{(\frac32)_s}\cos^{2s+1}\beta \; .
\end{eqnarray}

The eigenvalues (\ref{KKevenbranch}) and (\ref{KKoddbranch}) actually correspond to a unique tower of KK graviton masses. This is made apparent by introducing a new quantum number $n$ defined as 
\begin{equation} \label{nQN}
n =
\left\{
\begin{array}{lll}
2j + k  & ,                        & \textrm{for the first branch}  \\
2j + 1 + k   & ,                 & \textrm{for the second branch} \; .
 \end{array} \right.
\end{equation}
In terms of $(n,k)$, (\ref{KKevenbranch}) and (\ref{KKoddbranch}) can be combined into the single KK tower:
\begin{equation} \label{KKbranchG2}
 g^{-2}M_{n,k}^2=e^{-3\varphi}X^{3}n(n+6) + e^{-\varphi}X(e^{2\varphi}-e^{-2\varphi}X^2)k(k+5) \; ,
\end{equation} 
which is our final result. The quantum numbers range here as 
\begin{eqnarray} \label{QNranges}
n = 0, 1, 2 , \ldots \; , \qquad
k = 0 , 1 , \ldots , n \; .
\end{eqnarray}
Only $n$ ranges freely over the non-negative integers, due to its definition (\ref{nQN}) in terms of the non-negative but otherwise unconstrained integer $j$. The range of $k$ is limited to $k \leq n $ by (\ref{nQN}). At fixed $n$, the eigenvalue (\ref{KKbranchG2}) occurs with degeneracy
\begin{equation} \label{MassDeg}
D_{k,7} \equiv \textrm{dim} \, [k,0,0]_{\textrm{SO}(7)}   \; ,
\end{equation}
where, more generally, $D_{k,N}$ is the dimension of the symmetric traceless representation $[k,0, \ldots, 0]$ of SO$(N)$,
\begin{eqnarray} \label{kSONdim}
D_{k , N} &=  & { k+N-1   \choose k } - { k+N-3   \choose k -2 }  \\[5pt]
&=& \tfrac{1}{(N-2)!} \, (2k+N-2) (k+N-3) (k+N-4) \cdots (k+2)(k+1) \; , \nonumber
\end{eqnarray}
for $k \geq 2$ and
\begin{equation} \label{kSONdimk=0}
D_{0 , N} = 1 \; , \qquad
D_{1 , N} = N \; , \qquad  \textrm{for all $N= 2 , 3 \ldots$}
\end{equation}
It is also useful to note that
\begin{equation} \label{eq:DimRel}
D_{n,N-1} = D_{n,N} -  D_{n-1, \, N}   \; , \qquad \textrm{for all $n= 1 , 2 , \ldots \ $ and all $N= 2 , 3  \ldots$}
\end{equation}
The eigenfunctions (\ref{EigenFunceven}), (\ref{EigenFuncodd}) can be similarly combined into
\begin{equation} \label{KKbranchEigenFG2}
\cy_{n,k} = \cy_k \sin^{k}\beta \sum_{s=0}^{\left[ \frac{n-k}{2} \right]}  (-1)^s \binom{\left[ \frac{n-k}{2} \right]}{s} \frac{(\left[ \frac{n-k}{2} \right] +k+3+h_{n,k})_s}{(\frac{1}{2} + h_{n,k})_s} \cos^{2s+h_{n,k}}\beta \; ,
\end{equation}
where $[ \, ]$ means integer part and we define the symbol $h_{n,k}$ as
\begin{equation}\label{hnkSymbol}
h_{n,k} =  n-k - 2\left[ \frac{n-k}{2} \right] = 
\left\{
\begin{array}{lll}
	0  & ,                  & n-k \textrm{ even (for the first branch)}  \\
	1  & ,                  & n-k \textrm{ odd (for the second branch)} \; .
\end{array} \right.
\end{equation}

At fixed $n$ and $k$, the eigenfunctions (\ref{KKbranchEigenFG2}) span the $[k,0, 0]$ representation of $\textrm{SO}(7)_v$. Moreover, it can be checked that these eigenfunctions at fixed $n$ actually span the full symmetric traceless representation $[n,0, 0,0]$ of SO(8). In other words, the eigenfunctions (\ref{KKbranchEigenFG2}) turn out to be simply the SO(8) spherical harmonics of $S^7$, branched out into $\textrm{SO}(7)_v$ representations through
\begin{equation}
  \label{eq:SO7toSO6toU3branching}
  [n, 0, 0,0] \;
  \stackrel{\mathrm{SO}(7)_v}{\longrightarrow} \; 
   \sum_{k=0}^n \, [k ,0 ,0] 
    \; .
\end{equation}
This is consistent with the quantum number ranges (\ref{QNranges}). This is also compatible with the internal geometry (\ref{eq:metG2}) being topologically $S^7$: it can be continuously deformed into the round SO(8)--invariant geometry by setting $\varphi = \chi = 0$. These arguments suggest that the spectrum (\ref{KKbranchG2}), (\ref{KKbranchEigenFG2}) is in fact complete. Thus, the quantum number $n$ can be regarded as the Kaluza-Klein level, as it coincides with the unique integer that characterises the KK spectrum of the $\cN=8$ SO(8)-invariant Freund-Rubin solution: see {\it e.g.} table 9 of \cite{Duff:1986hr}.

\subsubsection{Massive gravitons with at least SU(4)$_{c}$ symmetry} \label{sec:SU4sec}

The SU(4)$_{c}$-invariant sector of SO(8)-gauged supergravity contains three pseudoscalars: $\chi$, $\zeta$, $\tilde{\zeta} $ in the notation of appendix \ref{sec:SU3sector}. In the Iwasawa parametrisation of the appendix, the SU(3)--invariant dilatons $\varphi$, $\phi$ become identified in terms of the pseudoscalars  via equation (\ref{SU4csector}). With the understanding that $\varphi$, $\phi$ depend on the independent fields $\chi$, $\zeta$, $\tilde{\zeta} $, the former can be conveniently used to parametrise the SU(4)$_{c}$-invariant sector, as the resulting expressions are more compact. The embedding of this sector into the $D=11$ warp factor and internal metric reads \cite{Bobev:2010ib}, in the notation of \cite{Larios:2019kbw},
\begin{equation} \label{eq:embedSU4in11d}
	e^{2A}	=e^{\frac43\phi+\varphi}\, L^2 \; , \qquad 
	d\bar{s}^2_{7} 	=g^{-2}L^{-2}\left[e^{-2\phi-\varphi}ds^2(\mathbb{CP}^3)+e^{-3\varphi}(d\psi+\sigma)^2\right]\,.
\end{equation}
Here, $ds^2(\mathbb{CP}^3)$ is the Fubini-Study metric on the complex projective space, $\sigma$ a one-form potential for the K\"ahler form on the latter, and $0 \leq \psi \leq 2\pi$ a coordinate on the Hopf fibre of $S^7$. The constant $g$ is again the coupling of SO(8) supergravity and $L$ is fixed through (\ref{eq:AdSLV}) in terms of the SU(4)$_{c}$-invariant potential $V$, given by (\ref{eq:scalarpotSU3inSO8}) with the identifications (\ref{SU4csector}). Away from the SO$(7)_c$-invariant locus, (2.39) of \cite{Larios:2019kbw}, where the symmetry is enhanced accordingly, the geometry (\ref{eq:embedSU4in11d}) is invariant under $\textrm{SU}(4)_c \times \textrm{U}(1)$, with U(1) generated by $\partial_\psi$. This U(1) is broken by the $D=11$ supergravity four-form. 

The $D=11$ embedding (\ref{eq:embedSU4in11d}) of the SU(4)$_{c}$-invariant sector is homogeneous: the warp factor depends only on the $D=4$ scalars and not on the $S^7$ coordinates, and the metric $d\bar{s}^2_{7}$ corresponds to a homogeneous stretching of the $S^7$ geometry along its Hopf fibre. Therefore, the differential equation (\ref{eq: GeneralPDE})  simplifies for this geometry as
\begin{equation}	\label{eq:opMth}
	\Big[\big(e^{3\varphi}-e^{2\phi+\varphi}\big)\partial_\psi^2+e^{2\phi+\varphi}\square_{S^7}\Big] \cy = -g^{-2} M^2 \, \cy \;,	
\end{equation}
with $\square_{S^7}$ the Laplacian on the round, Einstein metric on $S^7$. The solutions of (\ref{eq:opMth}) are accordingly given by the SO(8) spherical harmonics on $S^7$, branched out into representations of the $\textrm{SU}(4)_c \times \textrm{U}(1)$ symmetry group of (\ref{eq:embedSU4in11d}) and (\ref{eq:opMth}) via
\begin{equation} \label{eq:SO8intoSU4U1}
	[n,0,0,0]\xrightarrow[]{\text{SU(4)}_c\times\text{U(1)}}\sum_{r=0}^n \, [r,0,n-r]_{2r-n} \; ,
\end{equation}
with the subindex indicating the U(1) charge. More concretely, the $S^7$ spherical harmonics, in the  $[n,0,0,0]$ of SO(8), split according to (\ref{eq:SO8intoSU4U1}) as
\begin{equation} \label{eq:YSU4U1}
	\cy_{n,r}(z,\bar{z})=c_{a_1\dots a_r}{}^{b_1\dots b_{n-r}}\,z^{a_1}\dots z^{a_r}\,\bar{z}_{b_1}\dots\bar{z}_{b_{n-r}} \; , 
\end{equation}
for
\begin{eqnarray} \label{QNrangesSU4}
n = 0, 1, 2 , \ldots \; , \qquad
r = 0 , 1 , \ldots , n \; .
\end{eqnarray}
In (\ref{eq:YSU4U1}), $z^1=\mu^1+i\mu^2$, etc, are complexified embedding coordinates of $\mathbb{R}^8$ constrained as $\delta_{AB} \, \mu^A\mu^B=1$, with $A,B=1,\dots,8$, and $c_{a_1 \dots a_r}{}^{b_1 \dots b_{n-r}}$ is a constant tensor in the $[n-r,0,r]$ of SU(4). The functions (\ref{eq:YSU4U1}) obey
\begin{equation}
	\square_{S^7}\cy_{n,r}=-n(n+6)\cy_{n,r} \; , 	\qquad \qquad
	\partial_\psi^2\cy_{n,r}=-(n-2r)^2\cy_{n,r}\; ,
\end{equation}
and thus satisfy the differential equation (\ref{eq:opMth}) with eigenvalue
\begin{equation}	\label{eq: specSU4sector}
	g^{-2}M^2_{n , r}= e^{2\phi+\varphi}n(n+6) + \big(e^{3\varphi}-e^{2\phi+\varphi}\big)(n-2r)^2 \; .
\end{equation}
This occurs with multiplicity 
\begin{equation} \label{eq:SU4mult}
	d_{n,r}=\dim[r, 0 , n-r ]_{\text{SU(4)}}=\tfrac1{12}(n+3)(r+1)(r+2)(n-r+1)(n-r+2)\;.
\end{equation}

To summarise, the complete spectrum of the eigenvalue equation (\ref{eq:opMth}) is (\ref{eq: specSU4sector}), (\ref{eq:YSU4U1}), with the quantum numbers ranging as in (\ref{QNrangesSU4}). The eigenvalues (\ref{eq: specSU4sector}) have multiplicity (\ref{eq:SU4mult}) and the eigenfunctions (\ref{eq:YSU4U1}) are simply the $S^7$ spherical harmonics split into $\textrm{SU}(4)_c \times \textrm{U}(1)$ representations through (\ref{eq:SO8intoSU4U1}). The eigenvalues have been given in terms of $D=4$ scalars. The massive KK graviton spectra about $D=11$ AdS$_4$ solutions in this sector are obtained by fixing the $D=4$ scalars to the corresponding vevs. Like in the case discussed in section \ref{sec:G2sec}, the integer $n$ is identified with the KK level by an argument similar to that put forward below (\ref{eq:SO7toSO6toU3branching}).

\subsection{Type IIB} \label{sec:SpecIIB}

We now move on to compute the graviton spectrum about the AdS$_4$ solutions of type IIB supergravity recently obtained in \cite{Guarino:2019oct}. These geometries arise upon consistent uplift \cite{Inverso:2016eet} on an S-fold geometry of AdS$_4$ vacua of $D=4$ $\cN=8$ gauged supergravity with dyonic $[ \textrm{SO}(6) \times \textrm{SO}(1,1)] \ltimes\mathbb{R}^{12}$ gauging \cite{DallAgata:2011aa,Dall'Agata:2014ita} (see also \cite{Bobev:2019jbi}). The resulting type IIB uplifts correspond to limiting Janus-type solutions \cite{Bak:2003jk,DHoker:2006vfr,DHoker:2007zhm,DHoker:2007hhe}. As in section \ref{sec:Mtheory}, we will focus in this section on solutions that preserve at least SU(3) symmetry. These were classified in \cite{Guarino:2019oct}. We will compute the generic graviton spectra for arbitrary constant values of the SU(3)-invariant scalars of the $D=4$ supergravity. 

The type IIB geometries under consideration are of the form (\ref{eq:Mthbkg}) with $d=6$ and\footnote{We have conveniently rescaled the metric and warp factor with respect to  \cite{Guarino:2019oct}.} \cite{Guarino:2019oct}
\begin{equation} \label{eq:embedSU3inIIB}
	e^{2A} = \sqrt{Y}e^{\varphi}\, L^2 \; , \qquad 
	d\bar{s}^2_{6} = \frac{e^{-\varphi}}{\sqrt{Y}}\,g^{-2}L^{-2}\Big[\sqrt{Y}e^{-2\varphi}d\eta^2+\frac1{\sqrt{Y}}\big(ds^2(\mathbb{CP}^2)+Y(d\tau +\sigma^\prime)^2\big)\Big] \,.
\end{equation}
The geometry inside the last parenthesis extends globally over a topological $S^5$, with $ds^2(\mathbb{CP}^2)$ the Fubini-Study metric on the complex projective plane within $S^5$ and $0 \leq \tau < 2\pi$ the Hopf fibre angle. The local one-form $\sigma^\prime$ is a potential for the K\"ahler form on $\mathbb{CP}^2$. The sixth internal coordinate $\eta$ will be taken to be periodic, $\eta\sim\eta+T$, with $T$ a positive number. The ten-dimensional geometry (\ref{eq:embedSU3inIIB}) also depends on the SU(3)-invariant scalars of appendix \ref{sec:SU3sector} both explicitly and through the combination $Y$ defined in (\ref{scalDefs}). Finally, $g$ is the gauge coupling constant of the $D=4$ supergravity, and $L$ is fixed through (\ref{eq:AdSLV}) in terms of the scalar potential $V$ given by (\ref{eq:scalarpotSU3inSO6}) with $\chi = 0$. For general values of the scalars, the geometry (\ref{eq:embedSU3inIIB}) displays an isometry group $\textrm{SU}(3) \times \textrm{U}(1)_\tau \times \textrm{U}(1)_\eta$, with the U$(1)_\eta$ factor broken by the type IIB fluxes. In particular, the type IIB fields charged under the S-duality group SL$(2,\mathbb{R})$ undergo a monodromy transformation as $\eta$ crosses through different periods \cite{Inverso:2016eet}. The type IIB metric is neutral under S-duality and thus insensitive to this transformation. 

Like in section \ref{sec:SU4sec}, the type IIB embedding (\ref{eq:embedSU3inIIB}) is homogeneous. Accordingly, the differential equation (\ref{eq: GeneralPDE}) reduces for this geometry to
\begin{equation}	\label{eq:opMthIIB}
	\Big[e^{3\varphi}\partial_\eta^2+e^{\varphi}(1-Y)\partial_\tau^2+e^{\varphi}Y \, \square_{S^5}\Big] \cy = -g^{-2} M^2 \, \cy \;,	
\end{equation}
where $\square_{S^5}$ is the Laplacian on the round, Einstein metric on $S^5$. The complete set of eigenfunctions $\cy \equiv \cy_{\ell,p,j}$ that solve (\ref{eq:opMthIIB}) can be taken to satisfy
\begin{equation} \label{eq:IIBHarmonics}
	\square_{S^5}\cy_{\ell,p,j} = -\ell (\ell+4)\cy_{\ell,p,j} \; , 	\qquad
	\partial_\tau^2\cy_{\ell,p,j} = -(\ell-2p)^2\cy_{\ell,p,j} \; , \qquad
	\partial_\eta^2\cy_{\ell,p,j} = -\big(\tfrac{2\pi}{T}j\big)^2 \, \cy_{\ell ,p,j} \; ,
\end{equation}
for
\begin{eqnarray} \label{QNrangesSU3IIB}
\ell = 0, 1, 2 , \ldots \; , \qquad
p = 0 , 1 , \ldots , \ell \; , \qquad
j = 0, \pm 1, \pm 2, \ldots \; ,
\end{eqnarray}
with $\ell$ and $j$ unconstrained and $p$ constrained by $\ell$ through $p \leq \ell$. In other words, the eigenfunctions $\cy_{\ell,p,j}$ come in representations of $\textrm{SU}(3) \times \textrm{U}(1)_\tau  \times \textrm{U}(1)_\eta$, and are explicitly given by products of harmonics on the $S^1$ generated by $\partial_\eta$ and spherical harmonics $[\ell,0,0]_{\textrm{SO}(6)}$ on $S^5$ branched out into representations of $\textrm{SU}(3) \times \textrm{U}(1)_\tau$ via
\begin{equation} \label{eq:SO6intoSO6U1IIB}
	[\ell,0,0]\xrightarrow[]{\text{SU(3)} \times\text{U(1)}_\tau}\sum_{p=0}^\ell \, [p,\ell-p]_{\ell-2p} \; .
\end{equation}
Bringing (\ref{eq:IIBHarmonics}) to (\ref{eq:opMthIIB}), we find the eigenvalues
\begin{equation} \label{eq: specSU3sectorIIB}
	g^{-2}M^2_{\ell,p,j}=  e^{\varphi}Y \ell (\ell+4) +e^{\varphi}(1-Y)(\ell-2p)^2 + e^{3\varphi}\big(\tfrac{2\pi}{T}j\big)^2   \;,
\end{equation}
occurring with degeneracy
\begin{equation} \label{eq:SU3IIBmult}
d_{\ell,p, j}=
\left\{
\begin{array}{lll}
\dim[p,\ell-p]_{\text{SU(3)}} \; \; \; \, = \tfrac12 \, (p+1)(\ell-p+1)(\ell+2)  & ,                        & \textrm{if $j=0$}  \\
2 \dim[p,\ell-p]_{\text{SU(3)}} \; \, = (p+1)(\ell-p+1)(\ell+2)  & ,                 & \textrm{if $j \neq 0$} \; .
 \end{array} \right.
\end{equation}

In summary, the complete eigenvalue spectrum of equation (\ref{eq:opMthIIB}) is (\ref{eq: specSU3sectorIIB}) with the eigenfunctions $\cy_{\ell,p,j}$ described above and with the quantum numbers ranging as in (\ref{QNrangesSU3IIB}). The eigenvalues (\ref{eq: specSU3sectorIIB}) have multiplicity (\ref{eq:SU3IIBmult}), and have been given in terms of $D=4$ scalars. The massive KK graviton spectra about $D=11$ AdS$_4$ solutions in this sector are obtained by fixing the $D=4$ scalars to the corresponding vevs, as we will see next.

\subsection{Individual spectra in M-theory, type IIA and type IIB} \label{sec:IndvidualSpectra}

Using the results of sections \ref{sec:Mtheory} and \ref{sec:SpecIIB} as well as \cite{Duff:1986hr,Klebanov:2009kp,Pang:2017omp}, we can write down the KK graviton spectra about the AdS$_4$ solutions of the ten and eleven-dimensional supergravities that uplift from critical points with at least SU(3) symmetry of the three $D=4$ $\cN=8$ gauged supergravities that we are considering in this paper.

\baselineskip=16pt
\begin{table}[]
\centering
\resizebox{\textwidth}{!}{
\begin{tabular}{lll}
\hline
Solution                
& Mass                  
& Degeneracy  
\\ \hline \hline
$\cN=8 \ , \; \textrm{SO}(8) $              
& $L^2 M_n^2 = \tfrac14 n(n+6)$                                 
& $d_n = D_{n,8}$ 
\\[5pt] \hline
\multirow{2}{*}{$\cN=2 \ , \; \textrm{SU}(3) \times \textrm{U}(1)_c $ }
&$
L^2 M^2_{n,p,\ell,r} =\tfrac12n(n+6) -\tfrac13 \ell (\ell +4) -\tfrac19(\ell -2p)^2 $
&\multirow{2}{*}{$d_{n,p,\ell,r} = \tfrac12 (p+1) (\ell -p +1) (\ell+2)$}
\\ 
& $ \qquad \qquad \,\,\,\, +\tfrac1{18}[3( n-2r ) +4 (\ell-2p)]^2$
&
\\[5pt] \hline
$\cN=1 \ , \; \textrm{G}_2$
&$
L^2 M_{n,k}^2 = \tfrac{5}{8}n(n+6) - \tfrac{5}{12}k(k+5)
$
&$d_{n,k} = D_{k,7} $
\\[5pt] \hline
$\cN=0 \ , \; \textrm{SO}(7)_v$ 
&$
L^2 M_{n,k}^2 = \tfrac{3}{4}n(n+6) - \tfrac{3}{5}k(k+5)
$  
& $d_{n,k} = D_{k,7} $ 
\\[5pt] \hline
$\cN=0 \ , \; \textrm{SO}(7)_c$
&
$
L^2 M_{n}^2 = \tfrac{3}{10}n(n+6)
$
&
$d_{n} = D_{n,8} $ 
\\[5pt] \hline
$\cN=0 \ , \; \textrm{SU}(4)_c$  
&$
L^2 M_{n,r}^2 = \tfrac{3}{8}n(n+6) - \tfrac{3}{16}(n-2r)^2
$
& $d_{n,r} = \tfrac1{12}(n+3)(r+1)(r+2)(n-r+1)(n-r+2)$
\\[5pt] \hline
\end{tabular}
}
\caption{\footnotesize{The KK graviton spectra of AdS$_4$ solutions of $D=11$ supergravity that uplift from critical points of $D=4$ $\cN=8$ SO(8)-gauged supergravity with at least SU(3) symmetry. See (\ref{kSONdim}) for the notation $D_{k , N}$. The quantum numbers range as in (\ref{QNrangesSO8}).  
}\normalsize}
\label{eq:KKGravSpectraSols11D}
\end{table}

In M-theory, the spectrum above the AdS$_4$ solutions with at least G$_2$ symmetry and at least $\textrm{SU}(4)_c$ symmetry can be obtained by particularising (\ref{KKbranchG2}) and (\ref{eq: specSU4sector}), respectively, to the scalar vevs given in \cite{Larios:2019kbw}. We have brought these results to table \ref{eq:KKGravSpectraSols11D}. In order to exhaust the KK graviton spectra of AdS$_4$ solutions of $D=11$ supergravity that uplift from critical points of $D=4$ $\cN=8$ SO(8)-gauged supergravity with at least SU(3) symmetry, the table also includes the spectrum \cite{Duff:1986hr} about the $\cN=8$ Freund-Rubin solution \cite{Freund:1980xh} and the spectrum \cite{Klebanov:2009kp} about the $\textrm{SU}(3) \times \textrm{U}(1)_c$--invariant AdS$_4$ solution \cite{Warner:1983vz,Corrado:2001nv}. The latter is given as in \cite{Pang:2017omp}, with $n_\textrm{here} = n_\textrm{there}$, $r_\textrm{here} = r_\textrm{there}$, $p_\textrm{here} = p_\textrm{there}$ and $\ell_\textrm{here} = p_\textrm{there} +  q_\textrm{there}$. The corresponding multiplicites are also given in the table, and the quantum numbers range as 
\begin{eqnarray} \label{QNrangesSO8}
n = 0, 1, 2 , \ldots \; , \qquad
r,\,k = 0 , 1 , \ldots , n \; , \qquad 
\ell = p ,  \ldots , p+ r \; , \qquad 
p = 0 , 1 , \ldots , n-r  . \; 
\end{eqnarray}
The only quantum number that is free to range unrestricted over the non-negative integers is $n$, all the others being bound by it. This is consistent with the interpretation of $n$ as the SO(8)  KK level, see below (\ref{eq:SO7toSO6toU3branching}). At fixed KK level $n$, the degeneracy of the $\cN=8$ SO(8)--symmetric spectrum is broken into representations of the isometry group of the internal metric. This may be larger than the symmetry of each solution, as the fluxes will further break the isometry to the actual symmetry quoted in the table. Similarly, the eigenfunctions corresponding to each solution are simply the $S^7$ spherical harmonics branched out into the representations of the relevant group. 

For convenience, table \ref{eq:KKGravSpectraSolsIIA} imports from \cite{Pang:2017omp} the KK graviton spectra of AdS$_4$ solutions of massive IIA supergravity that uplift from critical points of $D=4$ $\cN=8$ dyonic ISO(7)-gauged supergravity with at least SU(3) symmetry. The table includes the squared masses in units of the corresponding AdS radius $L$, as well as the multiplicites. In this case, the quantum numbers' ranges are 
\begin{eqnarray} \label{QNrangesIIA}
k = 0, 1, 2 , \ldots \; , \qquad
\ell = 0 , 1 , \ldots , k \; , \qquad 
p = 0 , 1 , \ldots , \ell \; ,
\end{eqnarray}
with $k_\textrm{here} = n_\textrm{in \cite{Pang:2017omp} }$. Again, $k$ is the only quantum number that is unrestricted. For this reason, $k$ can be interpreted in this case as the SO(7) KK level.
The eigenfunctions are now the $S^6$ spherical harmonics split into representations of the internal isometry group. This again may be larger than the symmetry of each solution given in table \ref{eq:KKGravSpectraSolsIIA} because the fluxes may further break the isometry to the actual symmetry of the metric and fluxes.

\baselineskip=16pt
\begin{table}[]
\centering
\resizebox{\textwidth}{!}{
\begin{tabular}{lll}
\hline
Solution                
& Mass                  
& Degeneracy  
\\ \hline \hline
$\cN=2 \ , \; \textrm{SU}(3) \times \textrm{U}(1)_v $
&$
L^2 M_{k,\ell,p}^2 = \tfrac23  k  ( k + 5) -\tfrac13    \ell  ( \ell + 4)  + \tfrac19  (\ell - 2p)^2 $
&$d_{k,\ell,p} = \tfrac12 (p+1) (\ell -p +1) (\ell+2)$  
\\[5pt] \hline
$\cN=1 \ , \; \textrm{G}_2$
&  $
L^2 M_{k}^2 = \tfrac{5}{12}  k  (  k + 5) 
$
& $d_{k} = D_{k,7} $ 
\\[5pt] \hline
$\cN=1 \ , \; \textrm{SU}(3)$  
&$
L^2 M_{k,\ell,p}^2 = \tfrac56  k  ( k + 5) -\tfrac{5}{12}    \ell  ( \ell + 4)  - \tfrac{5}{36}  (\ell - 2p)^2
$ 
& $d_{k,\ell,p} = \tfrac12 (p+1) (\ell -p +1) (\ell+2)$  
\\[5pt] \hline
$\cN=0 \ , \; \textrm{SO}(7)_v$  
&$
L^2 M_{k}^2 = \tfrac{2}{5}  k  ( k + 5)
$
& $d_{k} = D_{k,7} $  
\\[5pt] \hline
$\cN=0 \ , \; \textrm{SO}(6)_v$ 
& $
L^2 M_{k,\ell}^2 =  k  ( k + 5) -\tfrac{3}{4}    \ell  ( \ell + 4)  
$
& $d_{\ell} = D_{\ell,6} $  
\\[5pt] \hline
$\cN=0 \ , \; \textrm{G}_2$  
&$
L^2 M_{k}^2 = \tfrac{1}{2}  k  ( k + 5) 
$
& $d_{k} = D_{k,7} $  
\\[5pt] \hline
\end{tabular}
}

\caption{\footnotesize{The KK graviton spectra of AdS$_4$ solutions of massive IIA supergravity that uplift from critical points of $D=4$ $\cN=8$ dyonic ISO(7)-gauged supergravity with at least SU(3) symmetry, taken from \cite{Pang:2017omp}. See (\ref{kSONdim}) for the notation $D_{k , N}$. The quantum numbers range as in (\ref{QNrangesIIA}).}\normalsize}
\label{eq:KKGravSpectraSolsIIA}
\end{table}

Finally, we turn to the spectrum of gravitons corresponding to the type IIB AdS$_4$ S-fold solutions that uplift from critical points with at least SU(3) symmetry \cite{Guarino:2019oct} of $D=4$ $\cN=8$ supergravity with $\big( \textrm{SO}(6) \times \textrm{SO}(1,1) \big) \ltimes \mathbb{R}^{12}$ gauging. These are found by bringing the corresponding vevs, collected in our conventions in table \ref{table:so6xso11criticalpoints} in appendix \ref{sec:SU3sector}, to equation (\ref{eq: specSU3sectorIIB}). The results are summarised in table \ref{eq:KKGravSpectraSolsIIB}. In order to derive the generic scalar-dependent spectra in section \ref{sec:SpecIIB}, we assumed that the S-fold direction $\eta$ is compactified to a $\textrm{U}(1)_\eta$ with period $T$. The KK graviton spectra are sensitive to this period. The eigenfunctions are products of $S^5$ harmonics, possibly branched out into $\textrm{SU}(3) \times \textrm{U}(1)_\tau$ representations,  and $\textrm{U}(1)_\eta$ harmonics. This $\textrm{U}(1)_\eta$ is broken by the IIB fluxes.

\subsection{Universality of traces} \label{sec:UniversalTraces}

When regarded as vacua of their corresponding $D=4$ $\cN=8$ gauged supergravities, the AdS solutions under consideration with at least SU(3) symmetry tend to exhibit the same mass spectrum of scalars, vectors and fermions within their $D=4$ supergravities. This is the case for all these solutions, except for the two $\cN=0$, SU(3)--invariant critical points of ISO(7) supergravity and the $\cN=0$, SU(3)--invariant critical locus of $\big( \textrm{SO}(6) \times \textrm{SO}(1,1) \big) \ltimes \mathbb{R}^{12}$ supergravity. The question that we would like to address in this section is whether this situation persists for higher KK modes. The spectrum of gravitons computed for these solutions in section \ref{sec:IndvidualSpectra} shows that this universality is indeed lost at higher KK levels: the KK gravitons do have completely different masses for all the solutions considered.

However, as we will now show, universality is still maintained, though in a milder form that is not apparent from the results of section \ref{sec:IndvidualSpectra}. It turns out that certain sums of KK graviton masses weighted with their multiplicities do remain universal. This is the case at least for solutions in the same or different $\cN=8$ gaugings with the same symmetry and whose spectra within the $D=4$ supergravity are the same. Specifically, if two AdS$_4$ solutions of $D=11$ supergravity or massive IIA uplift from critical points with the same supersymmetry $\cN \leq 8$, the same symmetry $G \supset \textrm{SU}(3)$ (possibly embedded differently into the gauge group) and the same spectrum within the $D=4$ $\cN=8$ SO(8) or ISO(7) supergravities, then there exist infinitely many discrete combinations $L^2 \, \textrm{tr} \, M^2_\n$, $ n=1,2,3, \ldots$, of graviton masses weighted with their multiplicities that are the same for both solutions. This statement was proven for the $\cN=2$ $\textrm{SU}(3) \times \textrm{U}(1)$-invariant solutions in \cite{Pang:2017omp}. Here we will extend that result to all other solutions with at least SU(3) symmetry in the SO(8) and ISO(7) gaugings, summarised in table \ref{tab:SolutionsinGaugings} of the introduction. As discussed in \cite{Pang:2017omp} and further in section \ref{sec:KKGravMassMat} below, the notation $L^2 \, \textrm{tr} \, M^2_\n$ relates to the fact that the combinations in question correspond to traces of the (infinite-dimensional) KK graviton mass matrix at fixed KK level $n$.

\baselineskip=16pt
\begin{table}[]
\centering
\resizebox{\textwidth}{!}{
\begin{tabular}{lll}
\hline
Solution                
& Mass                  
& Degeneracy  
\\ \hline \hline
$\cN=1 \ , \; \textrm{SU}(3)$
&$
L^{2}M^2_{\ell,p,j}= \tfrac56 \ell(\ell+4) - \tfrac{5}{36} (\ell-2p)^2 + \tfrac{5 \pi^2}{T^2} j^2  $
& $d_{\ell,p,j}$
\\[5pt] \hline
\rule{0pt}{1.02\normalbaselineskip}
$\cN=0 \ , \; \textrm{SO}(6)_v$
&  $
L^{2}M^2_{\ell,j}= \tfrac34 \ell (\ell +4) + \tfrac{6 \pi^2}{T^2}  j^2
$
& $ d_{\ell, j} =  (2 - \delta_{j0}) \,  D_{\ell,6}$
\\[5pt] \hline
\rule{0pt}{1.02\normalbaselineskip}
$\cN=0 \ , \; \textrm{SU}(3)$  
&$
L^{2}M^2_{\ell,j}= \tfrac34 \ell (\ell +4) + \tfrac{6 \pi^2}{T^2}  j^2
$ 
& $ d_{\ell, j} =  (2 - \delta_{j0}) \, D_{\ell,6}$
\\[5pt] \hline
\end{tabular}
}

\caption{\footnotesize{The KK graviton spectra of AdS$_4$ S-fold solutions of type IIB supergravity that uplift from critical points of $D=4$ $\cN=8$ $\big(\textrm{SO}(6)\times\textrm{SO}(1,1)\big)\ltimes\mathbb{R}^{12}$-gauged supergravity with at least SU(3) symmetry. See (\ref{kSONdim}) for the notation $D_{k , N}$ and (\ref{eq:SU3IIBmult}) for $d_{\ell,p, j}$. The quantum numbers range as in (\ref{QNrangesSU3IIB}).}\normalsize}
\label{eq:KKGravSpectraSolsIIB}
\end{table}

More concretely, for the M-theory solutions we define $L^2 \,  \textrm{tr} \, M^2_\n$ to be the sum of the squared masses in units of the corresponding AdS radius $L$, weighted with the corresponding multiplicity as given in table \ref{eq:KKGravSpectraSols11D}. The sum is taken at fixed KK level $n$ and over all other quantum numbers ranging as in (\ref{QNrangesSO8}). For example, using this prescription, one obtains for the $\cN=8$ SO(8) solution \cite{Pang:2017omp},
\begin{equation}
L^2 \, \textrm{tr} \, M^2_\n  = L^2 M_n^2 \, d_n = 14 \, D_{n-1 , 10} \; .
\end{equation}
In the last step, we have made use of the definition (\ref{kSONdim}) as a shorthand for the resulting 8th degree polynomial in $n$. Similarly, for the $\cN=2$ $\textrm{SU}(3) \times \textrm{U}(1)_c$ solution, we have \cite{Pang:2017omp}
\begin{equation}
L^2 \, \textrm{tr} \, M^2_\n  = L^2  \sum_{r=0}^n \sum_{p=0}^{n-r} \sum_{\ell=p}^{p+r} M^2_{n, p , \ell , r}  \, d_{n, p , \ell , r}  = \tfrac{56}{3} \, D_{n-1 , 10} \; .
\end{equation}
Proceeding similarly, we compute the quantities $L^2 \,  \textrm{tr} \, M^2_\n$, $n= 1 , 2, \ldots $,  for the KK graviton spectra summarised in table \ref{eq:KKGravSpectraSols11D} for $D=11$ AdS$_4$ solutions that uplift from critical points of $D=4$ $\cN=8$ SO(8) supergravity with at least SU(3) symmetry. We obtain:
\begin{equation} \label{eq:TrSols11D}
\textrm{
\begin{tabular}{llll}
$\cN=8 \ , \; \textrm{SO}(8) $ & : &
$L^2 \,  \textrm{tr} \, M^2_\n = 14 \, D_{n-1 , 10}  $ & ,  \\[10pt]
$\cN=2 \ , \; \textrm{SU}(3) \times \textrm{U}(1)_c $  & : &  
$
L^2 \,  \textrm{tr} \, M^2_\n = \tfrac{56}{3} \, D_{n-1 , 10}$ &,  \\[10pt] 
$\cN=1 \ , \; \textrm{G}_2$  & : &  
$
L^2 \,  \textrm{tr} \, M^2_\n = \tfrac{35}{2} \, D_{n-1 , 10}
$  & , \\[10pt] 
$\cN=0 \ , \; \textrm{SO}(7)_v$  & : &  
$
L^2 \,  \textrm{tr} \, M^2_\n = \tfrac{84}{5} \, D_{n-1 , 10}
$  & ,  \\[10pt] 
$\cN=0 \ , \; \textrm{SO}(7)_c$  & : &  
$
L^2 \,  \textrm{tr} \, M^2_\n = \tfrac{84}{5} \, D_{n-1 , 10}
$  & ,   \\[10pt] 
$\cN=0 \ , \; \textrm{SU}(4)_c$  & : &  
$
L^2 \,  \textrm{tr} \, M^2_\n = \tfrac{39}{2} \, D_{n-1 , 10}
$ & .  
\end{tabular}
}
\end{equation}
In particular, the two SO(7)-invariant solutions have their residual symmetry embedded differently into the SO(8) gauge group as SO$(7)_v$ and SO$(7)_c$. They have the same mass spectrum within $D=4$ $\cN=8$ SO(8) supergravity, according to table \ref{tab:SolutionsinGaugings}. Their KK graviton spectra are different, though, according to table \ref{eq:KKGravSpectraSols11D}. But as can be seen from equation (\ref{eq:TrSols11D}), the quantity $L^2 \, \textrm{tr} \, M^2_\n$ is the same for both solutions for all $n$. 

The quantities $L^2 \,  \textrm{tr} \, M^2_\k$ for the KK gravitons of massive IIA solutions with at least SU(3) symmetry that uplift from critical points of dyonic ISO(7) supergravity were computed similarly, for $k=1,2, \ldots$, in \cite{Pang:2017omp}:
\begin{equation} \label{eq:TrSolsIIA}
\textrm{
\begin{tabular}{llll}
$\cN=2 \ , \; \textrm{SU}(3) \times \textrm{U}(1)_v $  & : &  
$
L^2 \;  \textrm{tr} \, M_\k^2 = \tfrac{56}{3}  \, D_{k-1, \, 9}
$  & ,  
  \\[10pt]
$\cN=1 \ , \; \textrm{G}_2$  & : &  
$
L^2 \;  \textrm{tr} \, M_\k^2 = \tfrac{35}{2}  \, D_{k-1, \, 9}
$  & ,  \\[10pt]
$\cN=1 \ , \; \textrm{SU}(3)$  & : &  
$
L^2 \;  \textrm{tr} \, M_\k^2 = \tfrac{65}{3}  \, D_{k-1, \, 9}
$  & , 
 \\[10pt]
$\cN=0 \ , \; \textrm{SO}(7)_v$  & : &  
$
L^2 \;  \textrm{tr} \, M_\k^2 = \tfrac{84}{5}  \, D_{k-1, \, 9}
$  & , 
\\[10pt]
$\cN=0 \ , \; \textrm{SO}(6)_v$  & : &  
$
L^2 \;  \textrm{tr} \, M_\k^2 = \tfrac{39}{2}  \, D_{k-1, \, 9} 
$  & , 
\\[10pt]
$\cN=0 \ , \; \textrm{G}_2$  & : &  
$
L^2 \;  \textrm{tr} \, M_\k^2 = 21 \, D_{k-1, \, 9}
$  & .
%
%
\end{tabular}
}
\end{equation}
Here, we have again made use of the notation $D_{k,N}$ defined in (\ref{kSONdim}) as a shorthand for the degree-7 polynomial in $k$ that apparears in the r.h.s.'s. Now, recall from section \ref{sec:IndvidualSpectra} that $k$ and $n$ can respectively be regarded as the KK levels in massive IIA and $D=11$. At first KK level, the quantities $L^2 \;  \textrm{tr} \, M_{\neqone}^2$ in (\ref{eq:TrSols11D}) and $L^2 \;  \textrm{tr} \, M_\keqone^2$ in (\ref{eq:TrSolsIIA}) can be checked to match, by virtue of the first relation in (\ref{kSONdimk=0}), for solutions with the same symmetry group regardless of the embedding of the latter within the corresponding gauge group. For example, for the $D=11$ $\textrm{SU}(3) \times \textrm{U}(1)_c$ solution \cite{Corrado:2001nv} and the massive IIA $\textrm{SU}(3) \times \textrm{U}(1)_v$ solution \cite{Guarino:2015jca}, $ [L^2 \;  \textrm{tr} \, M_{\1}^2]_{\textrm{$11$D}} = [L^2 \;  \textrm{tr} \, M_{\1}^2]_{\textrm{IIA}} = \frac{56}{3}$, at $n=k=1$, as already noted in \cite{Pang:2017omp}. Inspection of (\ref{eq:TrSols11D}) and (\ref{eq:TrSolsIIA}) confirms that similar matches occur at KK level one, $n=k=1$,  for the $D=11$ and massive IIA solutions with common (super)symmetry $\cN=1$, G$_2$, and $\cN=0$, SO(7), and $\cN=0$, $\textrm{SU}(4) \sim \textrm{SO}(6)$.

Further, there is still matching at higher KK levels $n >1$ in $D=11$ and $k >1$ in massive IIA, provided a prescription is adopted to relate $n$ and $k$. An argument will be given in section \ref{sec:KKGravMassMat} but, for now, these two quantum numbers can be thought of as being related as in (\ref{eq:SO7toSO6toU3branching}), so that the $D=11$ KK level $n$ formally contains all IIA KK levels $k = 0 , 1 , \ldots , n$. Using this prescription, it follows from (\ref{eq:TrSols11D}) and (\ref{eq:TrSolsIIA}) that 
\begin{eqnarray} \label{tracerelation1}
\sum_{k=0}^n \,  [L^2 \;  \textrm{tr} \, M^2_{\k} ]_\textrm{IIA} \,=    [ L^2 \; \textrm{tr} \, M^2_{\n} ]_\textrm{11D}  \; , \qquad n= 0, 1 , 2 , \ldots  \; ,
\end{eqnarray}
for all the solutions that we are considering with the same symmetry and supersymmetry in massive IIA and $D=11$. Here, $L^2 \;  \textrm{tr} \, M^2_{\0} \equiv 0$ corresponds to the massless graviton, for both the $D=11$ and type IIA cases, as well as for the IIB cases below. From (\ref{tracerelation1}) it immediately follows that 
\begin{eqnarray} \label{tracerelation2}
\sum_{n=0}^m \sum_{k=0}^n  \,  [L^2 \;  \textrm{tr} \, M^2_{\k} ]_\textrm{IIA} \,=  \sum_{n=0}^m  [ L^2 \; \textrm{tr} \, M^2_{\n} ]_\textrm{11D}  \; , \qquad m = 0, 1 , 2 , \ldots  \; ,
\end{eqnarray}
again for all solutions with the same (super)symmetry. The sums in (\ref{tracerelation2}) obviously run over repeated number of states, both in IIA and in $D=11$. In (\ref{tracerelation1}), there are no repeated $D=11$ states on the r.h.s., but the sum in the l.h.s. does run as well over repeated states in IIA. These overcounting issues can be avoided by subtracting two adjacent KK levels in $D=11$: formally, the difference between KK levels $n$ and $n-1$ in $D=11$ contains the same number of states as KK level $k=n$ in massive IIA. Using the identity (\ref{eq:DimRel}), it follows from (\ref{eq:TrSols11D}), (\ref{eq:TrSolsIIA}) that
\begin{eqnarray} \label{tracerelation3}
[L^2 \;  \textrm{tr} \, M^2_{\n} ]_\textrm{IIA} \,= \, [ L^2 \; \textrm{tr} \, M^2_{\n} ]_\textrm{11D} - [ L^2 \;  \textrm{tr} \, M^2_{(n-1)} ]_\textrm{11D}  \; , \qquad n= 1 , 2 , \ldots  \; , 
\end{eqnarray}
for solutions with the same (super)symmetry. This relation was already shown to hold in \cite{Pang:2017omp} for the $\cN=2$ $\textrm{SU}(3) \times \textrm{U}(1)$ invariant solutions. Here, we have extended this result to all other AdS solutions in the SU(3)-invariant sectors of SO(8) and ISO(7) gauged supergravities with the same symmetry and supersymmetry.

The situation is similar, though slightly different, for the type IIB AdS$_4$ S-fold solutions that uplift from  $D=4$ $\cN=8$ $\big(\textrm{SO}(6)\times\textrm{SO}(1,1)\big)\ltimes\mathbb{R}^{12}$-gauged supergravity. According to table \ref{tab:SolutionsinGaugings}, this supergravity also has critical points with the same symmetry $G \supset \textrm{SU}(3)$ and supersymmetry as other critical points of the SO(8) and ISO(7) gauging: $\cN=0$ SO(6), $\cN=1$ SU(3) and $\cN=0$ SU(3). The former two have the same spectrum within their corresponding $D=4$ supergravities, while the latter does not. For this reason, we will only be interested in the former two vacua. Both for the $\cN=1$ SU(3) and the $\cN=0$ $\textrm{SO}(6) \sim \textrm{SU}(4)$ solutions there are combinations, $[L^2 \,  \textrm{tr} \, \tilde{M}^2_\n]_{\textrm{IIB}}$, of the eigenvalues in table \ref{eq:KKGravSpectraSolsIIB} that match the quantities $[L^2 \;  \textrm{tr} \, M^2_{\k} ]_\textrm{IIA}$ and $[L^2 \;  \textrm{tr} \, M^2_{\n} ]_\textrm{11D}$ for the solutions with the same symmetry for a certain choice of the period $T$. The tilde in $[L^2 \,  \textrm{tr} \, \tilde{M}^2_\n]_{\textrm{IIB}}$ is taken to signify that, in this case, the combinations also involve subtraction of eigenvalues. More concretely, consider the following quantities for the type IIB solutions with the quantum numbers fixed as indicated:
\begin{equation} \label{eq:TrSolsIIB}
\textrm{
\begin{tabular}{lll}
$\cN=1 \ , \; \textrm{SU}(3) $  & : &  
$
L^2 \;  \textrm{tr} \, \tilde{M}_\1^2 \equiv L^{2} \big[ \sum_{p=0}^{\ell} \, M^2_{\ell,p,j} d_{\ell,p,j} \big] |_{\ell=1, j = 0 }   $
 \\
& & $ \qquad\qquad  -L^{2} \big[  M^2_{\ell,p,j} d_{\ell,p,j} \big] |_{\ell= p = 0  ,  j = -1 }  -L^{2} \big[  M^2_{\ell,p,j} d_{\ell,p,j} \big] |_{\ell= p = 0  ,  j = +1 }  
$ ,  
  \\[15pt]
$\cN=0 \ , \; \textrm{SO}(6)_v$  & : &  
$
L^2 \;  \textrm{tr} \, \tilde{M}_\1^2 \equiv L^{2}   \big[ M^2_{\ell,j} d_{\ell,j} \big] |_{\ell=1, j = 0 }   $
 \\
& & $ \qquad\qquad\; \;\;   -L^{2} \big[  M^2_{\ell,j} d_{\ell,j} \big] |_{\ell=  0  ,  j = -1 }  -L^{2} \big[  M^2_{\ell,j} d_{\ell,j} \big] |_{\ell= 0  ,  j = +1 }  
$ .
%
%
\end{tabular}
}
\end{equation}

These quantities involve sums of mass eigenvalues, weighted with their degeneracies as given in table \ref{eq:KKGravSpectraSolsIIB}, and affected by a $+$ or a $-$ sign depending on whether $j=0$ or $j \neq 0$. Plugging in the expressions given in the table, the quantity $L^2 \;  \textrm{tr} \, \tilde{M}_\1^2$ for the $\cN=1$, SU(3) solution evaluates to $\frac{65}{3}$ if $T=2\pi$, matching the quantity $L^2 \;  \textrm{tr} \, M_\keqone^2$ for its counterpart type IIA solution at KK level $k=1$, given in (\ref{eq:TrSolsIIA}). Similarly, $L^2 \;  \textrm{tr} \, \tilde{M}_\1^2$ for the $\cN=0$, $\textrm{SO}(6)_v$ solution evaluated using the expressions given in table \ref{eq:KKGravSpectraSolsIIB} gives $\frac{39}{2}$ for $T=2\pi$. This again matches the quantity $L^2 \;  \textrm{tr} \, M_\keqone^2$ at KK level $k=1$ given in (\ref{eq:TrSolsIIA}) for the $\cN=0$, $\textrm{SO}(6)_v$ solution of massive IIA. It also matches $L^2 \;  \textrm{tr} \, M_\neqone^2$ at KK level $n=1$ given in (\ref{eq:TrSols11D}) for the $D=11$ $\cN=0$, $\textrm{SU}(4)_c$ solution. Although it is not as clear cut in the type IIB case, it will be argued in section \ref{sec:Discussion} that the states that enter the sums in (\ref{eq:TrSolsIIB}) also belong to KK level $m=1$ in an SL(8)-covariant sense. The formal analytic continuation $j^\prime = i j $, with $i^2 = -1$, removes the minus signs in (\ref{eq:TrSolsIIB}). Under this analytic continuation, relations similar to (\ref{tracerelation1}) and (\ref{tracerelation2}) relate these formal sums at higher KK levels for these type IIB solutions to their $D=11$ and type IIA counterparts. We will return to this point in section \ref{sec:Discussion}. In the next section, we will turn to compute the KK graviton spectrum of a different type IIB S-fold solution of particular interest.


\section{Graviton spectrum on the $\cn=4\,\,$ $\sofour$ type IIB S-fold } \label{sec:IIBSO4Spectrum}

The spin-2 spectrum about the AdS$_4$ solution \cite{Inverso:2016eet} of type IIB supergravity that uplifts on a six-dimensional S-fold from the $\cN=4$ SO(4)-invariant critical point \cite{Gallerati:2014xra} of $D=4$ $\cN=8$ $[ \textrm{SO}(6) \times \textrm{SO}(1,1)] \ltimes\mathbb{R}^{12}$--gauged supergravity \cite{DallAgata:2011aa,Dall'Agata:2014ita} can be computed as in section \ref{sec: SpecMth}. This solution has concrete field theory duals for specific choices of the period $T$ of the S-fold coordinate $\eta$ \cite{Assel:2018vtq} (see also \cite{Assel:2011xz,Assel:2012cj,Garozzo:2018kra,Garozzo:2019hbf,Garozzo:2019ejm}). For this reason, it is particularly interesting to go in some detail about the corresponding spin-2 spectrum. 

The relevant ten-dimensional geometry is (\ref{eq:Mthbkg}) with $d=6$ and \cite{Inverso:2016eet}
{\setlength\arraycolsep{2pt}
\begin{eqnarray} \label{eq:embedSO4inIIB}
	d\bar{s}^2_{6} & = & g^{-2}L^{-2}\Big[  d\eta^{2}+ \frac{dr^{2}}{1-r^{2}}+\frac{r^{2}}{1+2r^{2}} \, ds^2 (S_1^2)+\frac{1-r^{2}}{3-2r^{2}} \, ds^2 (S_2^2)\Big] \, , \nonumber  \\[8pt] 
	e^{2A} & = & L^2 \Big[(1+2r^{2})(3-2r^{2})\Big]^{1/4} \; ,
\end{eqnarray}
}with $L^2 =\frac12 g^{-2}$. The coordinate $r$ ranges as $0 \leq r \leq 1$ and $\eta$ is taken to be periodic, $\eta\sim\eta+T$, for some $T >0$. Also, $ds^2 (S_i^2)$, $i=1,2$, is the round, Einstein metric on each of two spheres $S^2_i$. These are rotated by the $\textrm{SO}(4) = \textrm{SO}(3)_1 \times \textrm{SO}(3)_2$ isometry of the geometry (\ref{eq:embedSO4inIIB}). This $\textrm{SO}(4)$ isometry is also respected by the type IIB forms and is thus a symmetry, in fact the R-symmetry, of the full $\cN=4$ ten-dimensional solution. In contrast, the $\textrm{U}(1)_\eta$ isometry generated by $\partial_\eta$ is broken by the supergravity forms. Topologically, the metric (\ref{eq:embedSO4inIIB}) extends over $S^5 \times S^1_\eta$ \cite{Inverso:2016eet}, with the $S^5$ directions corresponding to $r$ and $S^2_i$, $i=1,2$. 

Next, we plug the geometry (\ref{eq:embedSO4inIIB}) into the PDE (\ref{eq: GeneralPDE}). It is natural to separate the eigenfunction $\cy$ following the $\textrm{SO}(3)_1 \times \textrm{SO}(3)_2 \times \textrm{U}(1)_\eta$ isometry as 
\begin{equation} \label{eq:sepSO4}
\cy=f(r) \, \cy_\1 \, \cy_\2 \, \cy_\eta \; , 
\end{equation}
where $\cy_\1$, $\cy_\2$ and $\cy_\eta$ are the spherical harmonics on $S^2_1$, $S^2_2$ and $S^1_\eta$,
\begin{equation} \label{eq:IIBHarmonicsSO4sol}
	\square_{S^2_1}\cy_\1 = -\ell_1 (\ell_1+1)\cy_\1 \; , 	\qquad
	\square_{S^2_2}\cy_\2 = -\ell_2 (\ell_2+1)\cy_\2 \; , \qquad
	\partial_\eta^2\cy_\eta = -\big(\tfrac{2\pi}{T}j\big)^2 \, \cy_\eta \; ,
\end{equation}
with
\begin{eqnarray} \label{QNrangesSO4IIB1}
\ell_1 = 0, 1, 2 , \ldots \; , \qquad
\ell_2 = 0, 1, 2 , \ldots \; , \qquad
j = 0, \pm 1, \pm 2, \ldots 
\end{eqnarray}

The separation of variables (\ref{eq:sepSO4}) turns the PDE (\ref{eq: GeneralPDE}) into the following ODE in $r$:
\begin{equation} \label{eq:rEqIIB}
(1-r^{2}) f^{\prime\prime} +\Big(\frac{2}{r}-5r\Big) f^\prime+\Big[2M^{2}L^2 -\frac{4\pi^2}{T^2 } \, j^2 -\frac{1+2r^{2}}{r^{2}}\ell_{1}(\ell_{1}+1)-\frac{3-2r^{2}}{1-r^{2}}\ell_{2}(\ell_{2}+1)\Big] f=0 ,
\end{equation}
with a prime denoting derivative with respect to $r$. Finally, the change of variables 
\begin{equation} \label{eq:VarChangeSO4IIB}
r^{2}=u\,,\qquad f=u^{\ell_{1}/2}(1-u)^{\ell_{2}/2} \, H (u)  \,,
\end{equation}
where now $0 \leq u \leq 1$, brings the ODE (\ref{eq:rEqIIB}) into hypergeometric form (\ref{eq:HGeq}) with
\begin{equation} \label{eq:HGparamsIIB}
    a_\pm = \tfrac12 \big( \ell_{1}+\ell_{2} + 2 \big) \pm \tfrac12 \sqrt{4+ 2M^{2} L^2  -2\ell_{1}(\ell_{1}+1)-2\ell_{2}(\ell_{2}+1) - \tfrac{4\pi^2}{T^2} \, j^2 }  \; , 
    \hspace{0.5cm}
    c = \ell_{1}+\tfrac{3}{2} \, .
\end{equation}

The two linearly independent solutions to (\ref{eq:HGeq}) are written in (\ref{LinIndepeHGSols}). In the present case, the second solution therein blows up at $u=0$ since $\ell_1 \geq 0$ by (\ref{QNrangesSO4IIB1}), and is therefore excluded. The first solution in (\ref{LinIndepeHGSols}) is  regular at $u=0$ for all values of the parameters (\ref{eq:HGparamsIIB}), and is also regular at $u=1$ if $a_- = - h$ for a non-negative integer $h$. For this choice, the eigenfunction becomes a polynomial in $u$. Defining a new quantum number
\begin{equation} \label{eq:hintermsofl}
\ell=2h+\ell_{1}+\ell_{2}\,,
\end{equation}
the graviton masses then follow from (\ref{eq:HGparamsIIB}) as 
\begin{equation} \label{eq:GravMassSO4IIB}
L^2 M_{\ell,\ell_{1},\ell_{2}, j }^{2}= \tfrac{1}{2} \ell(\ell+4)+\ell_{1}(\ell_{1}+1)+\ell_{2}(\ell_{2}+1) + \tfrac{2\pi^2}{T^2} \, j^2  \,,
\end{equation}
with the quantum numbers now ranging as 
\begin{equation} \label{eq:QNrangesSO4IIB}
\ell = 0 ,1, 2 , \ldots \; , \qquad 
\ell_{1}=0,1,\ldots,\ell\,,\qquad
\ell_{2}=0,1,\ldots,\ell\,,\qquad
j = 0, \pm 1, \pm 2, \ldots \; , 
\end{equation}
by (\ref{QNrangesSO4IIB1}), (\ref{eq:hintermsofl}) and the fact that $h \geq 0$. The number of KK gravitons with mass (\ref{eq:GravMassSO4IIB}) is

\begin{equation} \label{eq:SU3IIBmultSO4}
d_{\ell,\ell_1,\ell_2, j}=
\left\{
\begin{array}{lll}
 \; \, (2\ell_{1}+1)(2\ell_{2}+1)   & ,                        & \textrm{if $j=0$}  \\
2 (2\ell_{1}+1)(2\ell_{2}+1)  & ,                 & \textrm{if $j \neq 0$} \; .
 \end{array} \right.
\end{equation}

To summarise, the KK gravitons about the $\cN=4$ SO(4)-invariant AdS$_4$ S-fold solution of type IIB supergravity reported in \cite{Inverso:2016eet} have masses (\ref{eq:GravMassSO4IIB}), with quantum numbers ranging as in (\ref{eq:QNrangesSO4IIB}) and degeneracy (\ref{eq:SU3IIBmultSO4}). The corresponding eigenfunctions are given by (\ref{eq:sepSO4}) with (\ref{eq:VarChangeSO4IIB}) and the first hypergeometric function  in (\ref{LinIndepeHGSols}) for $H(u)$, which now becomes a polynomial. More precisely, the eigenfunctions (\ref{eq:sepSO4}) are products of $S^5$ spherical harmonics branched out in $\textrm{SO}(4) = \textrm{SO}(3)_1 \times \textrm{SO}(3)_2$ representations, and harmonics on $S^1_\eta$. A few of these modes for low values of the quantum numbers have been tabulated in table \ref{tab:spectrumN=4Sfold}. The table includes the dimension $\Delta$ of the corresponding operators in the dual field theory \cite{Assel:2018vtq}, where
\begin{equation} \label{eq:M2Delta}
M^{2} L^2 =\Delta(\Delta-3) \; .
\end{equation}
The field theory is defined on a stack of D3-branes wrapped on $S^1_\eta$. The form of the dual single-trace spin-2 operators can be inferred from the supergravity eigenfunctions. They correspond to bound states of the energy-momentum tensor $T_{\mu\nu}$, the six scalars $X_\1^a$, $X_\2^a$, $a=1,2,3$, corresponding to the directions transverse to the D3-branes, and a complex coordinate $Z$ on the D3-brane. In the table, all these have been promoted to superfields. The trace in the adjoint of the gauge group is understood.

	 \begin{table}[]
\centering

\resizebox*{\textwidth}{!}
{

\begin{tabular}{|c|l|c|c|c|c|c|}
\hline
$m$                & $(\ell,\ell_1,\ell_2,|j|)$                  & $d_{\ell,\ell_1,\ell_2,j}$   &  $L^2 M^2_{\ell,\ell_1,\ell_2,j}$    & $\Delta_{\ell,\ell_1,\ell_2,j}$                                                       & Dual operator                                                          & Short?       \\ \hline \hline
0                  & $ (0,0,0,0)$                                  & 1   & 0               & 3                                               & $\mathcal{T}^{(0)}_{\alpha\beta}\vert_{s=2}$                      & $\checkmark$ 

\\ \hline
\multirow{2}{*}{1} & $ (1,1,0,0)\,,\,(1,0,1,0)  $                                  & 6   & $\frac{9}{2}  $  &   $\frac{3}{2} \left(\sqrt{3}+1\right)$                                                  
 &  $\mathcal{T}^{(0)}_{\alpha\beta}\mathcal{R}\mathcal{X}^a_{(1)}\vert_{s=2}\, , \, S_{(1)} \leftrightarrow S_{(2)}$                                  &              
                   \\ \cline{2-5} \cline{6-7} 
                   & $(0,0,0,1) $ & 2   & $\frac{2 \pi ^2}{T^2} $  & $\frac{1}{2} \left(\sqrt{9+\frac{8 \pi ^2}{T^2} }+3\right)$                                                                      & $\mathcal{T}^{(0)}_{\alpha\beta} \mathcal{Z}\vert_{s=2}$, c.c.             & $ $ 
                   
                   \\ \hline
\multirow{7}{*}{2} & $(2,0,0,0) $                                  & 1   & $6$  & $\frac{1}{2} \left(\sqrt{33}+3\right)$  &$\mathcal{T}^{(0)}_{\alpha\beta}(1-2 \mathcal{R}^2)\vert_{s=2}$                                                      &   $ $           
\\ \cline{2-5} \cline{6-7} 
                   & $ (2,1,1,0) $ & 9   & $10$  & $5$                                     
  &$\mathcal{T}^{(0)}_{\alpha\beta}\mathcal{R}(1-\mathcal{R}^2)^{1/2}
  \mathcal{X}_{(1)}^a \mathcal{X}_{(2)}^b \vert_{s=2}$                                    
                     &   $\checkmark $        
                        \\ \cline{2-5} \cline{6-7} 
                   &$ (2,2,0,0)\,,\,(2,0,2,0) $ & 10   & $12$  & $\frac{1}{2} \left(\sqrt{57}+3\right)$                                                                      & $\mathcal{T}^{(0)}_{\alpha\beta}\mathcal{R}^2
 ( \mathcal{X}_{(1)}^{(a} \mathcal{X}_{(1)}^{b)}-\frac{1}{3}\mathcal{R}^2 \delta^{ab}) \vert_{s=2} \, , \, S_{(1)} \leftrightarrow S_{(2)}$                                       & $ $
\\ \cline{2-5} \cline{6-7} 
                   & $(1,1,0,1)\,,\,(1,0,1,1) $                                  & 12   & $\frac{2 \pi ^2}{T^2}+\frac{9}{2}$  
  & $\frac{1}{2} \left(\sqrt{27+ \frac{8 \pi ^2}{T^2}}+3\right)$                                  &$\mathcal{T}^{(0)}_{\alpha\beta}\mathcal{R}\mathcal{X}_{(1)}^{a}\mathcal{Z}\vert_{s=2}$ ,c.c. , $S_{(1)} \leftrightarrow S_{(2)} $                                     &              
\\ \cline{2-5} \cline{6-7} 
                   & $(0,0,0,2) $                                  & 2   & $\frac{8 \pi ^2}{T^2}$  
& $\frac{1}{2} \left(\sqrt{9+\frac{32 \pi ^2}{T^2}}+3\right)$                                     & $\mathcal{T}^{(0)}_{\alpha\beta}\mathcal{Z}^2\vert_{s=2}$ , c.c.                                     &              
                   \\ \cline{2-5} \cline{6-7} 
                   & $(0,0,0,0) $ Redundant                                  & 1   & $0$  & $3$  &                                                              $\mathcal{T}^{(0)}_{\alpha\beta}\vert_{s=2}$                                     & $\checkmark$

                   \\ \hline
\multirow{14}{*}{3} & $ (3,1,0,0)\,,\,(3,0,1,0) $                                  & 6   & $\frac{25}{2} $           & $\frac{1}{2} \left(\sqrt{59}+3\right)$            
& $\mathcal{T}^{(0)}_{\alpha\beta}
\mathcal{R}(1-\frac{8}{5}\mathcal{R}^2)\mathcal{X}^a_{(1)}\vert_{s=2}
, , \, S_{(1)} \leftrightarrow S_{(2)} $                                               &              \\ \cline{2-5} \cline{6-7} 
                   & $(3,2,1,0)\,,\,(3,1,2,0)$ & 30   & $\frac{37}{2}$ & $\frac{1}{2} \left(\sqrt{83}+3\right)$                                    
&$\mathcal{T}^{(0)}_{\alpha\beta}\mathcal{R}^2
(\mathcal{X}_{(1)}^{(a}\mathcal{X}_{(1)}^{b)}-\frac{1}{3}\mathcal{R}^2\delta^{ab})\mathcal{X}_{(2)}^{c}\vert_{s=2}\, , \, S_{(1)} \leftrightarrow S_{(2)} $                                                                &              \\ \cline{2-5} \cline{6-7} 
                   & $(3,3,0,0)\,,\,(3,0,3,0)$ & 14   & $\frac{45}{2}$ & $\frac{3}{2} \left(\sqrt{11}+1\right)$                                                                    &  $\mathcal{T}^{(0)}_{\alpha\beta}\mathcal{R}^3
(\mathcal{X}_{(1)}^{(a}\mathcal{X}_{(1)}^{b}\mathcal{X}_{(1)}^{c)}-\text{traces})\vert_{s=2}\, , \, S_{(1)} \leftrightarrow S_{(2)} $  &              
 \\ \cline{2-5} \cline{6-7} 
& $(2,0,0,1)$                                 & 2   
& $\frac{2 \pi ^2}{T^2}+6$              
& $\frac{1}{2} \left(\sqrt{33+\frac{8\pi^2}{T^2}}+3\right)$                                               &$\mathcal{T}^{(0)}_{\alpha\beta}(1-2 \mathcal{R}^2) \mathcal{Z}\vert_{s=2}$ , c.c.                         &             
 \\ \cline{2-5} \cline{6-7} 
                   & $(2,1,1,1)$                      & 18  
& $\frac{2 \pi ^2}{T^2}+10$              
& $\frac{1}{2} \left(\sqrt{49+\frac{8\pi^2}{T^2}}+3\right)$                                                    & $\mathcal{T}^{(0)}_{\alpha\beta}\mathcal{R}(1- \mathcal{R}^2)^{1/2} \mathcal{X}^a_{(1)} \mathcal{X}^b_{(2)} \mathcal{Z}\vert_{s=2}$ , c.c.  & $ $ 
\\ \cline{2-5} \cline{6-7} 
& $(2,2,0,1)\,,\,(2,0,2,1)$ & 20  
& $\frac{2 \pi ^2}{T^2}+12$  
& $\frac{1}{2} \left(\sqrt{57+\frac{8\pi^2}{T^2}}+3\right)$                                                   & $\mathcal{T}^{(0)}_{\alpha\beta}\mathcal{R}^2 (\mathcal{X}_{(1)}^{(a}\mathcal{X}_{(1)}^{b)}-\frac{1}{3}\mathcal{R}^2\delta^{ab})\mathcal{Z}\vert_{s=2}$ , c.c.  ,  $S_{(1)} \leftrightarrow S_{(2)} $
 &      
\\ \cline{2-5} \cline{6-7} 
&$(1,1,0,2)\,,\,(1,0,1,2)$ & 12  
& $\frac{8 \pi ^2}{T^2}+\frac{9}{2}$  
& $\frac{1}{2} \left(\sqrt{27+\frac{32\pi^2}{T^2}}+3\right)$                                                   & $\mathcal{T}^{(0)}_{\alpha\beta}\mathcal{R}\mathcal{X}_{(1)}^{(a)} \mathcal{Z}^2\vert_{s=2}$ , c.c.  ,  $S_{(1)} \leftrightarrow S_{(2)} $ &
\\ \cline{2-5} \cline{6-7} 
& $(0,0,0,3)$ & 2  
& $\frac{18 \pi ^2}{T^2}$  
& $\frac{3}{2} \left(\sqrt{1+\frac{8\pi^2}{T^2}}+1\right)$ &                                                  $\mathcal{T}^{(0)}_{\alpha\beta} \mathcal{Z}^3\vert_{s=2}$ , c.c.                                               & 
\\ \cline{2-5} \cline{6-7} 
                   & $(1,1,0,0)\,,\,(1,0,1,0)$ Redundant & 12 & $\frac{9}{2}  $  &   $\frac{3}{2} \left(\sqrt{3}+1\right)$                                                 
 & $\mathcal{T}^{(0)}_{\alpha\beta}\mathcal{R}\mathcal{X}^a_{(1)}\vert_{s=2}\, , \, S_{(1)} \leftrightarrow S_{(2)}$                                  &  
\\ \cline{2-5} \cline{6-7} 
& $(0,0,0,1)$ Redundant & 4   
& $\frac{2 \pi ^2}{T^2}$  
& $\frac{1}{2} \left(\sqrt{9+\frac{8 \pi ^2}{T^2}}+3\right)$                                                                      & $\mathcal{T}^{(0)}_{\alpha\beta} \mathcal{Z}\vert_{s=2}$, c.c.             & $ $ 
                           \\ \hline
\end{tabular}
}

\caption{\footnotesize{The spin-2 spectrum of the $\cN=4$ S-fold solution. We have employed the notation $\mathcal{R}^2=\delta_{ab} \mathcal{X}^a_{(1)}\mathcal{X}^b_{(1)}$ and $S_{(1)} \leftrightarrow S_{(2)}$ means exchange of the labels (1) and (2), and simultaneously $\mathcal{R}^2 \leftrightarrow 1-\mathcal{R}^2$. The spectrum is organised in SL(8) KK levels (see section \ref{sec:Discussion}) $m=0 , 1 , \ldots$, and this leads to some redundant states as discussed in that section. The dual operators are single-trace. The $\textrm{tr}$ symbol has been omitted.
}\normalsize}
\label{tab:spectrumN=4Sfold}
\end{table}

The full KK spectrum about this $\cN=4$ solution must lie in representations of OSp$(4|4)$. A classification of these representations can be found in, for example, \cite{Cordova:2016emh}. The massless graviton should belong to a short graviton multiplet, $A_2$ in the notation of \cite{Cordova:2016emh}, with $\ell_1 = \ell_2 = 0$. This multiplet contains gravitini and vectors in the fundamental and the adjoint, respectively, of the R-symmetry group SO(4). It also contains spin-$1/2$ fields in the fundamental of SO(4) and two singlet scalars, see {\it e.g.} equation (5.55) of \cite{Cordova:2016emh}. By the results of \cite{Gauntlett:2007ma,Malek:2017njj,Cassani:2019vcl}, a consistent truncation of type IIB supergravity must exist at the full non-linear level on this S-fold solution to the pure $D=4$ $\cN=4$ gauged supergravity with the field content just described. The SO(4)-singlet scalars in this $D=4$ $\cN=4$ supergravity parametrise an $\textrm{SL}(2, \mathbb{R})/\textrm{SO}(2)$ non-linear sigma model. The numerator group is inherited from the S-duality of type IIB supergravity.

The supergroup OSp$(4|4)$ also admits massive short representations. Some gravitons in the spectrum can be identified to belong to such representations, particularly $A_2$ in the notation of \cite{Cordova:2016emh}. Spin-2 states in these short multiplets arise as $Q^4$ descendants of the superconformal primary therein, and thus saturate the unitarity bound 
\begin{equation} \label{eq:IIBShortDeltas}
\Delta=\ell_{1}+\ell_{2}+3\, .
\end{equation}
Via (\ref{eq:M2Delta}), these states have masses 
\begin{equation} \label{eq:IIBShortSpectrum}
L^2 M^{2} =(\ell_{1}+\ell_{2})(\ell_{1}+\ell_{2}+3) \; .
\end{equation} 
In the spectrum (\ref{eq:GravMassSO4IIB}), short states of this type do indeed arise for any period $T$ and all $\ell_1 = 0 , 1, 2 , \ldots$, whenever $j=0$ and the other quantum numbers $\ell, \ell_1, \ell_2$ are related through $2 \ell_1 = 2 \ell_2 = \ell$ (so that $h=0$ in (\ref{eq:hintermsofl}) and $\ell$ is even). From (\ref{eq:GravMassSO4IIB}) and (\ref{eq:M2Delta}), the masses and conformal dimensions of states with these quantum numbers are
\begin{equation}
L^2 M_{\ell_1}^2 = 2 \ell_{1} (2 \ell_{1}+3)  \; , \qquad 
\Delta_{\ell_1} = 2\ell_{1}+3 \; , \qquad \ell_1 = 0 , 1 , 2 , \ldots \; , 
\end{equation}
which are indeed of the form (\ref{eq:IIBShortSpectrum}), (\ref{eq:IIBShortDeltas}) and thus short. 

All other gravitons belong to long multiplets. It can be checked that, away from the shortening relations among the quantum numbers, their $\Delta$'s computed through (\ref{eq:M2Delta}) from (\ref{eq:IIBShortSpectrum}) are always above the unitarity bound (\ref{eq:IIBShortDeltas}). We conclude, however, with the following tantalising observation. Consider the analytical continuation of the quantum number $j$ into $j^\prime = i j$, with $i^2=-1$, as at the end of section \ref{sec:UniversalTraces}, and fix the S-fold coordinate period to $T= 2\pi$. Then, it follows from (\ref{eq:GravMassSO4IIB}), that states with $j^\prime = \pm 1$, and $\ell_{2}=\ell_{1}+1$, and $\ell = 2 \ell_1 +1$ (so that $h=0$ in (\ref{eq:hintermsofl}) and $\ell$ is odd) are also short with conformal dimensions $\Delta_{\ell_1} = 2 (\ell_{1}+2)$ and masses $L^2 M_{\ell_1}^2 = 4 \ell_{1} (\ell_{1}+2) $, for all $\ell_1 = 0 ,  1 , 2 , \ldots$


\section{Kaluza-Klein graviton mass matrix} \label{sec:KKGravMassMat}


We now switch gears to obtain a covariant expression for the infinite-dimensional KK graviton mass matrix and its associated trace formulae.

\subsection{The mass matrix} \label{eq:MassMatCov}

We would like to determine the KK graviton mass matrix corresponding to string/M-theory AdS$_4$ solutions that uplift, at least, on the relevant spheres from the SO(8) and ISO(7) gaugings. In appendix A of \cite{Pang:2017omp}, an SO(7)-covariant mass matrix was derived for KK gravitons about solutions that uplift from the ISO(7) gauging. Here, we would like to extend those results into a mass matrix that is formally SL(8) covariant, in agreement with the formal, manifest covariance that the SO(8) and ISO(7) gauged supergravitites take on using the embedding tensor formalism \cite{deWit:2007mt} particularised to gaugings contained in $\textrm{SL}(8) \subset \textrm{E}_{7(7)}$. 

In order to do this, we start by assuming that the mass eigenfunctions, 
\begin{equation}
\cy^{A_1\dots A_m}=\mu^{(A_1}\dots\mu^{A_m)} \; , \qquad m = 0 , 1 , 2 , \ldots \; , 
\end{equation}
are symmetric polynomials of the $\mathbb{R}^8$ coordinates $\mu^A$, $A=1 , \ldots , 8$. The latter are formally in the fundamental of SL(8) and constrained as
\begin{equation} \label{eq:SninR8}
\theta_{AB} \, \mu^A \mu^B = 1 \; , 
\end{equation}
with $\theta_{AB} = \delta_{AB}$ for the SO(8) gauging and $\theta_{AB} = \textrm{diag} ( 1, 1 , 1, 1, 1, 1, 1, 0)$ for the ISO(7) gauging. We then pose the KK graviton mass equation (\ref{eq: GeneralPDE}) for the consistent embedding metrics on the $S^7$ and $S^6$ as given in \cite{Varela:2015ywx,Guarino:2015jca}. With these assumptions, we follow similar steps to those in \cite{Pang:2017omp} to transform the PDE (\ref{eq: GeneralPDE}) into an algebraic eigenvalue problem by reading off an infinite-dimensional block-diagonal mass matrix,
\begin{equation} \label{eq:KKMassMat}
\cM^2 = \textrm{diag} \big( M_\0^2 , M_\1^2 , M_\2^2 , \ldots , M_\m^2 ,  \ldots \big) \; .
\end{equation}
Now, each block is an $\textrm{SL}(8)$--covariant square matrix of size 
\begin{equation}
\textrm{dim} \, M_\m^2 \equiv [m,0,0,0,0,0,0]_{\textrm{SL}(8)} = { m+7   \choose m } \; .
\end{equation}
These blocks are given explicitly by the following expressions: for $m=0$,
\begin{equation} \label{eq:CovMassMatm=0}
	M^2_\0=0\;,
\end{equation}
for $m=1$,
\begin{equation} \label{eq:CovMassMatm=1}
	(M^2_\1)_A{}^B=-g^2\mathcal{M}^{\mathbb{M}\mathbb{N}} \, \Theta_{\mathbb{M}}{}^B{}_C\, \Theta_{\mathbb{N}}{}^C{}_A\;,
\end{equation}
for $m=2$,
\begin{equation} \label{eq:CovMassMatm=2}
	(M^2_\2)_{A_1A_2}{}^{B_1B_2}=-2g^2\mathcal{M}^{\mathbb{M}\mathbb{N}}\Big[\Theta_{\mathbb{M}}{}^{(B_1\vert}{}_C\;\Theta_{\mathbb{N}}{}^C{}_{(A_1}\delta_{A_2)}\!{}^{\vert B_2)}+
	\Theta_{\mathbb{M}}{}^{(B_1}{}_{(A_1}\;\Theta_{\mathbb{N}}{}^{B_2)}{}_{A_2)}\Big]\;,
\end{equation}
and for $m \geq 3$,
\begin{equation} \label{eq:CovMassMatmgeq3}
	\begin{aligned}
		(M^2_\m)_{A_1\dots A_m}{}^{B_1\dots B_m}=-m\,g^2\mathcal{M}^{\mathbb{M}\mathbb{N}}
		&\Big[\Theta_{\mathbb{M}}{}^{(B_1\vert}{}_C\;\Theta_{\mathbb{N}}{}^C{}_{(A_1}\delta_{A_2}\!{}^{\vert B_2}\dots \delta_{A_m)}\!{}^{\vert B_m)}		\\
		&+(m-1)\Theta_{\mathbb{M}}{}^{(B_1}{}_{(A_1}\;\Theta_{\mathbb{N}}{}^{B_2}{}_{A_2}\delta_{A_3}\!{}^{ B_3}\dots\delta_{A_m)}\!{}^{ B_m)}\Big]\; .
	\end{aligned}
\end{equation}
Compared to \cite{Pang:2017omp}, these expressions involve no trace removal within same-level indices. We have also restored the embedding tensors, $\Theta_\mathbb{M}{}^A{}_B$, with $\mathbb{M} = (_{[AB]} , ^{[AB]})$ a fundamental E$_{7(7)}$ index, and
\begin{equation} \label{eq:SL8}
	\Theta_{[AB]}{}^C{}_D=2 \, \delta^C_{[A}\theta_{B]D}\,,
	\qquad
	\Theta^{[AB]}{}^C{}_D= 2 \, \delta_D^{[A} \xi^{B]C} \; ,
\end{equation}
where $\theta_{AB}$ was defined below (\ref{eq:SninR8}), and $\xi^{AB} = 0$ for the SO(8) gauging while $\xi^{AB} = \textrm{diag} ( 0, 0 , 0, 0, 0, 0, 0, m/g)$ for ISO(7). Finally, $g$ here and in (\ref{eq:CovMassMatm=1})--(\ref{eq:CovMassMatmgeq3}) is the electric gauge coupling, $m$ the magnetic coupling, and $\mathcal{M}^{\mathbb{M}\mathbb{N}}$ is the inverse $\cN=8$ scalar matrix. Like the bosonic mass matrices of $D=4$ $\cN=8$ gauged supergravity (see  \cite{Trigiante:2016mnt}), the KK graviton mass matrices (\ref{eq:CovMassMatm=0})--(\ref{eq:CovMassMatmgeq3}) are quadratic in the $D=4$ embedding tensor.

Since we have refrained ourselves from removing traces on same-level indices of the KK graviton mass matrices (\ref{eq:CovMassMatm=0})--(\ref{eq:CovMassMatmgeq3}), the latter are manifestly SL(8)--covariant. We can think of these as blocks in the diagonal of the infinite dimensional graviton mass matrix (\ref{eq:KKMassMat}). The integer $m$ can be thought of as an SL(8) KK level. Proceeding like this, though, the price one pays for the SO(8) gauging is that the spectrum at fixed SL(8) KK level $m \geq 0$ contains repeated physical modes: it includes modes of all SO(8) KK levels $n$ (as defined below (\ref{eq:SO7toSO6toU3branching})) such that $n=m-2s$ according to 
\begin{equation}
  \label{eq:SL8toSO8}
  [m, 0, 0,0, 0, 0,0]_{\textrm{SL}(8)} \;
  \stackrel{\mathrm{SO}(8)}{\longrightarrow} \; 
   \sum_{s=0}^{\left[ \tfrac{m}2 \right]} \, [m-2s ,0 ,0,0]_{\textrm{SO}(8)} \; .
\end{equation}
For the ISO(7) gauging, there is an even larger overcounting. Every SL(8) level $m \geq 0$ formally contains the SO(8) levels $n$ specified in (\ref{eq:SL8toSO8}), and each of these, in turn, includes all SO(7) levels $k = 0 , 1 , \ldots , n$ by (\ref{eq:SO7toSO6toU3branching}). The repeated states can be projected out following (\ref{eq:SL8toSO8}) and (\ref{eq:SO7toSO6toU3branching}), leaving only physical modes. We also remark that it is the full embedding tensor for the dyonic ISO(7) gauging, including the magnetic contributions from $\xi^{AB}$, that enters (\ref{eq:CovMassMatm=0})--(\ref{eq:CovMassMatmgeq3}) for this gauging.

Taking into account this overcounting, we have verified up to SL(8) KK level $m=3$ that, particularising the KK graviton mass matrices (\ref{eq:CovMassMatm=0})--(\ref{eq:CovMassMatmgeq3}) to the corresponding critical points of the SO(8) and ISO(7) gaugings, their eigenvalues correctly reproduce the spectra given in tables \ref{eq:KKGravSpectraSols11D} and \ref{eq:KKGravSpectraSolsIIA}. Note that, in order to obtain matching, $g^2$ here has to be traded for $L^2$ there by making use of (\ref{eq:Lagrangian}) with the appropriate scalar potential (\ref{eq:scalarpotSU3inSO8}) or (\ref{eq:scalarpotSU3inISO7}). See section \ref{sec:Discussion} for a discussion of the case corresponding to the $[ \textrm{SO}(6) \times \textrm{SO}(1,1)] \ltimes\mathbb{R}^{12}$ gauging.

\subsection{Mass matrix traces}

By (\ref{eq:SL8toSO8}), the first SL(8) KK level $m=1$ contains only the first SO(8) KK level $n=1$  (in our conventions, $\bm{8} \rightarrow \bm{8}_v$). For the ISO(7) gauging, the first SL(8) KK level contains the SO(7) KK levels $k=0$, which has zero mass by (\ref{eq:CovMassMatm=0}), and $k=1$. At SL(8) KK level  $m=1$, the eigenvalues of the SL(8)-covariant mass matrix (\ref{eq:CovMassMatm=1}) thus reproduce the first KK-level eigenvalues with no overcounting for the SO(8) gauging. For the ISO(7) gauging, the first SL(8) level also reproduces the $k=1$ eigenvalues, together with an extra zero eigenvalue corresponding to $k=0$. The trace of the SL(8) level $m=1$ mass matrix (\ref{eq:CovMassMatm=1}),
\begin{equation} \label{eq:CovMassMatm=1Trace}
	\textrm{tr} \, M^2_\1=-g^2\mathcal{M}^{\mathbb{M}\mathbb{N}} \, \Theta_{\mathbb{M}}{}^A{}_B\, \Theta_{\mathbb{N}}{}^B{}_A\;,
\end{equation}
must thus reproduce the KK level-one traces discussed in section \ref{sec:UniversalTraces}, which indeed it does. Particularising (\ref{eq:CovMassMatm=1Trace}) to each specific critical point with at least SU(3) symmetry of the SO(8) and ISO(7) gaugings, making use of the relevant embedding tensors, and again trading $g^2$ for $L^2$, all the r.h.s.'s of (\ref{eq:TrSols11D}) and (\ref{eq:TrSolsIIA}) with $n=1$ and $k=1$ are reproduced. For example, using the appropriate embedding tensors and vevs, we find that (\ref{eq:CovMassMatm=1Trace}) evaluates to $\frac{56}{3}$, both for the $\textrm{SU}(3) \times \textrm{U}(1)_c$ point of the SO(8) gauging and for the $\textrm{SU}(3) \times \textrm{U}(1)_v$ point of the ISO(7) gauging, once that $g^2$ is replaced with the relevant $L^2$. The trace relation (\ref{tracerelation2}) is a direct consequence of (\ref{eq:CovMassMatm=1Trace}) and the overcounting feature mentioned in section \ref{eq:MassMatCov}. 

In order to check consistency, it is also useful to evaluate (\ref{eq:CovMassMatm=1Trace}) at generic invariant loci of the $D=4$ $\cN=8$ scalar manifold. For example, on the SU(3)-invariant sector of the SO(8) gauging, (\ref{eq:CovMassMatm=1Trace}) reduces to
\begin{equation} \label{eq:TraceSU3v1}
\text{tr} \, M^2_\1 =
2 g^2 e^{-3 \varphi -2 \phi } 
\Big[X^3 e^{2 \phi }+6 e^{2 \varphi } X \left(Y^2+Z^2\right)+3 (X+4 Y) e^{4 \varphi +2 \phi }+6 X e^{2 \varphi +4 \phi }\Big] \; ,
\end{equation}
where the shorthand notations (\ref{scalDefs}) have been employed. Further restricting (\ref{eq:TraceSU3v1}) to the G$_2$-invariant locus (\ref{eq:G2sector}), we find
\begin{equation} \label{eq:TraceG2sector}
 \text{tr} \, M^2_\1 = 14 g^2 \, (e^{-3\varphi} X^3 + 3e^{\varphi}X) \; , 
\end{equation}
in agreement with (\ref{KKbranchG2}). Alternatively, particularising (\ref{eq:TraceSU3v1}) to the SU$(4)_c$-invariant sector through (\ref{SU4csector}), we get
\begin{equation} \label{eq:TraceSU4csector}
	\text{tr} \, M_\1^2 = 8 g^{2} \, \big(e^{3\varphi}+6e^{2\phi+\varphi}\big) \; .
\end{equation}
This expression can be retrieved from (\ref{eq: specSU4sector}). Similarly, for the dyonic ISO(7) gauging and SU(3)-invariant scalars, (\ref{eq:CovMassMatm=1Trace}) reduces to
\begin{equation} \label{eq:TraceISO7}
\text{tr} \, M^2_\1 =
6 g^2  \big[e^{ \varphi } (X+4 Y)+2 X e^{2 \phi-\varphi  } \big] \; , 
\end{equation}
in agreement with (2.34) of \cite{Pang:2017omp}. 

For the SO(8) and dyonic ISO(7) gaugings, only the electric embedding tensor $\theta_{AB}$ actually participates in the trace formula (\ref{eq:CovMassMatm=1Trace}), while its magnetic counterpart $\xi^{AB}$ drops out. In order to see this, we expand (\ref{eq:CovMassMatm=1Trace}) using the generic form (\ref{eq:SL8}) of the embedding tensor, to obtain 
\begin{equation} \label{eq:CovMassMatExp}
\textrm{tr} \, M^2_\1 =-g^2 \left(
\mathcal{M}^{ACDE} \,\theta_{CD} \, \theta_{EA} 
+ \mathcal{M}_{ACDE} \,\xi^{CD} \, \xi^{EA} 
- 2 \mathcal{M}^{CD}\,_{CF} \,\theta_{DA} \, \xi^{FA} 
\right)
\; .
\end{equation}
For the SO(8) gauging, the statement is immediate because $\xi^{AB}=0$, and only the first term is non-vanishing . For the dyonic ISO(7) gauging, the only relevant term in (\ref{eq:CovMassMatExp}) is again the first one, because ${\cal M}_{8888} = 0$ and $\theta \xi = 0 $.


\section{Discussion} \label{sec:Discussion}


We have verified that all solutions with the same residual supersymmetry, the same bosonic symmetry containing SU(3) and the same spectrum within the SO(8) \cite{deWit:1982ig}, ISO(7) \cite{Guarino:2015qaa} and $\big(\textrm{SO}(6)\times\textrm{SO}(1,1)\big)\ltimes\mathbb{R}^{12}$ \cite{DallAgata:2011aa} gaugings all have different spectra of KK gravitons. The universality of the lowest KK level spectra is thus lost at higher KK levels. However, following \cite{Pang:2017omp}, we have found that this universality still persists up the different KK graviton towers in a weaker sense. Certain sums of KK graviton masses remain the same for all such solutions. We emphasise that this is the case if the gauged supergravity solutions with the same (super)symmetries also had the same spectrum at lowest KK level, namely, within their corresponding $\cN=8$ supergravities. 

For example, the $\cN=0$ SU(3)-invariant points of dyonic ISO(7) \cite{Guarino:2015qaa} and $\big(\textrm{SO}(6)\times\textrm{SO}(1,1)\big)\ltimes\mathbb{R}^{12}$ supergravities \cite{Guarino:2019oct} do not have the same spectrum within the $D=4$ supergravities. The relevant sums of KK graviton masses are also different. A similar observation holds for the U(1)-invariant $\cN=1$ points of ISO(7) supergravity recently reported in \cite{Guarino:2019snw}. Although they have the same residual symmetry, these points have different spectra within ISO(7) supergravity. Using the formula (\ref{eq:CovMassMatm=1Trace}), we have verified at KK level one that the relevant KK graviton mass sums are also different for these two solutions. It is also worth stressing that the trace universality property works for solutions that check out the above requirements independently of how the common residual symmetry group is embedded into the corresponding gauge group and ultimately E$_{7(7)}$. This was already noted in \cite{Pang:2017omp}, where these observations were made for the $\cN=2$ $\textrm{SU}(3) \times \textrm{U}(1)$-invariant points of SO(8) and ISO(7) supergravity. In the former case, the solution is embedded as $\textrm{SU}(3) \times \textrm{U}(1)_c$ and in the latter as $\textrm{SU}(3) \times \textrm{U}(1)_v$.

The relevant sums of KK graviton masses have been argued to be related to traces of the KK graviton mass matrix at fixed KK level. We have provided an SL(8)-covariant expression for the latter in section \ref{sec:KKGravMassMat}. This mass matrix has qualitatively the same form as the bosonic mass matrices of $D=4$ $\cN=8$ gauged supergravity in that it is quadratic in the embedding tensor and depends on the E$_{7(7)}/\textrm{SU}(8)$ $\cN=8$ (inverse) scalar matrix ${\cal M}^{\mathbb{M}\mathbb{N}}$. The SL(8)-covariant KK graviton mass matrix (\ref{eq:CovMassMatm=0})--(\ref{eq:CovMassMatmgeq3}) reproduces the KK graviton spectra of tables \ref{eq:KKGravSpectraSols11D} and \ref{eq:KKGravSpectraSolsIIA} for the uplifts of SO(8) and ISO(7) gauged supergravity critical points, with redundancies introduced by the SL(8) representations (versus branchings of SO(8) and SO(7) representations, as explained in section \ref{eq:MassMatCov}). Interestingly, the mass matrix (\ref{eq:CovMassMatm=0})--(\ref{eq:CovMassMatmgeq3}) also reproduces the KK graviton spectrum of table \ref{eq:KKGravSpectraSolsIIB} and equation (\ref{eq:GravMassSO4IIB}) for type IIB S-folds with period $T=2\pi$ that uplift from vacua of the $\big(\textrm{SO}(6)\times\textrm{SO}(1,1)\big)\ltimes\mathbb{R}^{12}$ gauging, {\it provided} the U$(1)_\eta$ quantum number $j$ is analytically continued as $j^\prime = i j $, with $i^2 = -1$. The origin of this analytic continuation can be put down to the fact that the SL(8)-covariant graviton mass matrix formula (\ref{eq:CovMassMatm=0})--(\ref{eq:CovMassMatmgeq3}) actually sees the compactified U$(1)_\eta$ as the non-compact SO$(1,1)$ factor in the $D=4$ gauge group $\big(\textrm{SO}(6)\times\textrm{SO}(1,1)\big)\ltimes\mathbb{R}^{12}$. From a IIB perspective, this factor is associated to a hyperboloid uplift \cite{Inverso:2016eet}. In any case, the (analytically continued) spectra of the type IIB S-folds can be also organised in SL(8) KK levels $m= 0 , 1 , \ldots$ through the branching 
\begin{equation}
  \label{eq:SL8toSO6xSO11}
  [m, 0, 0,0, 0, 0,0]_{\textrm{SL}(8)} \;
  \stackrel{\mathrm{SO}(6)\times\mathrm{SO}(1,1)}{\longrightarrow} \; 
   \sum_{s=0}^{\big[ \tfrac{m}2\big]}\sum_{\ell=0}^{m-2s}\sum_{p=0}^{m-2s-\ell} \, [\ell ,0,0]_{m-2s-\ell-2p} \; .
\end{equation}
This approach contains redundant states that can be projected out as discussed below (\ref{eq:SL8toSO8}). It would be interesting to determine if (\ref{eq:CovMassMatm=0})--(\ref{eq:CovMassMatmgeq3}) can be modified in such a way that the KK graviton spectra of vacua of the SO(8) and ISO(7) gauging are still obtained, and the physical spectra of the compactified S-folds is recovered as well.

The SL(8)-covariant KK graviton mass matrix should have an E$_{7(7)}$-covariant extension. From our analysis, it is not immediate to deduce what that extension should be. The trace formula (\ref{eq:CovMassMatm=1Trace}) does admit a naive straightforward E$_{7(7)}$-covariant extension, though:
\begin{equation} \label{eq:E7CovMassMatm=1Trace}
	\textrm{tr} \, M^2_\1=-g^2\, \kappa_{\alpha \beta}  \, \mathcal{M}^{\mathbb{M}\mathbb{N}} \, \Theta_{\mathbb{M}}{}^\alpha \, \Theta_{\mathbb{N}}{}^\beta\;,
\end{equation}
with $\kappa_{\alpha \beta}$ the Killing-Cartan form of E$_{7(7)}$. This formula reduces to (\ref{eq:CovMassMatm=1Trace}) for gaugings contained in SL(8). It would be interesting to test (\ref{eq:E7CovMassMatm=1Trace}) for gaugings not contained in SL(8) that descend from higher dimensions. An example of such gaugings, which has Minkowski vacua though, is given in \cite{Malek:2017cle}.


\section*{Acknowledgements}


KD, PN and OV are supported by the NSF grant PHY-1720364. GL is supported by an FPI-UAM predoctoral fellowship. GL and OV are partially sup\-por\-ted by grants SEV-2016-0597 and PGC2018-095976-B-C21 from MCIU/AEI/FEDER, UE.

\appendix

\addtocontents{toc}{\setcounter{tocdepth}{1}}

\section{SU(3) invariance in $D=4$ $\cN=8$ gauged supergravity} \label{sec:SU3sector}

We find it useful to collect here some facts about the SU(3)-invariant sector of the three different gaugings of $D=4$ $\cN=8$ supergravity considered in the main text. The field content is of course the same for all the gaugings considered but the interactions differ.

\begin{table}[t]
\centering\footnotesize
\ra{2}
\begin{tabular}{cc|cccccc|cc}
\Xhline{1pt}
	$\cN$	&	$G_0$		&	$\chi$	&	$e^{-\varphi}$	&	$e^{-\phi}$	&	$a$	&	$\zeta$		&	$\tilde{\zeta}$	&	$V_0$	& $L^2M^2$	\\
\Xhline{1pt}
	1		&	SU(3)		&	0		&	$\frac{\sqrt5m}{3g}$	&	$\sqrt{\frac56}$	&	0	&$\frac1{\sqrt3}$	&	$\frac1{\sqrt3}$	&	$-\frac{648 g^3}{25 \sqrt{5}m}$
				&	$\big(0,0,4-\sqrt{6},4-\sqrt{6},\sqrt{6}+4,\sqrt{6}+4\big)$		\\
	0		&	SO$(6)_v$		&	0		&$\frac{m}{\sqrt2 g}$	&		1		&	0	&		0		&		0		&	$-\frac{8 \sqrt{2} g^3}{m}$		
				&	$\big(0,0,6,6,-\frac{3}{4},-\frac{3}{4}\big)$	\\
	0		&	SU(3)		&	$\chi$	&$\frac{m}{\sqrt2 g}$&	$(1-a^2)^{1/4}$	&	$a$	&		0		&		0		&	$-\frac{8 \sqrt{2} g^3}{m}$
				&	$\big(0,0,6,6,\frac{27 g^2 \chi ^2}{2 m^2}-\frac{3}{4},\frac{27 g^2 \chi ^2}{2 m^2}-\frac{3}{4}\big)$	\\[1mm]

\Xhline{1pt}
\end{tabular}
\caption{\small{All critical loci of $D=4$ $\cN=8$ $[ \textrm{SO}(6) \times \textrm{SO}(1,1)] \ltimes\mathbb{R}^{12}$-gauged supergravity with at least SU(3) invariance. All of these are AdS. For each point we give the residual supersymmetry $\cN$ and bosonic symmetry $G_0$ within the full $\cN=8$ theory, their location in the parametrisation that we are using and the cosmological constant $V_0$ and the masses of the scalars in units of the AdS radius. The $\cN=0$ SO$(6)_v$ vacuum is the $\chi = a = 0 $ point of the $\cN=0$ SU(3) critical locus.}}
\label{table:so6xso11criticalpoints}
\end{table}

The SU(3)-invariant sector contains three scalars, $\varphi$, $\phi$, $a$, and three pseudoscalars $\chi$, $\zeta$, $\tilde{\zeta}$. All these are coordinates on a submanifold 
\begin{eqnarray} 
\label{ScalManN=2}
\frac{\textrm{SU}(1,1)}{\textrm{U}(1)} \times  \frac{\textrm{SU}(2,1)}{\textrm{SU}(2) \times \textrm{U}(1)} \,   
\end{eqnarray}
of E$_{7(7)}/$SU(8), with the first factor parametrised by $(\varphi, \chi)$, and the second by $(\phi, a , \zeta , \tilde{\zeta})$. The Lagrangian in this sector is
\begin{eqnarray}	\label{eq:Lagrangian}
{\cal L} &=&  R \, \textrm{vol}_4 + \tfrac{3}{2} ( d\varphi )^2  +  \tfrac{3}{2} e^{2 \varphi} \, ( d\chi )^2 +  2(D\phi )^2 + \tfrac{1}{2} \, e^{4 \phi} \,   \big( Da +  \tfrac{1}{2}  ( \zeta D \tilde{\zeta} - \tilde{\zeta} D \zeta  ) \big)^2    \\[5pt]
&&  +  \tfrac{1}{2} \, e^{2 \phi} \,  ( D\zeta )^2    +  \tfrac{1}{2} \, e^{2 \phi} \,   (D\tilde{\zeta} )^2 +  \tfrac{1}{2} \, \mathcal{I}_{\Lambda\Sigma} \, H_{\2}^{\Lambda} \wedge  * H_{\2}^{\Sigma}  + \tfrac{1}{2} \, \mathcal{R}_{\Lambda\Sigma} \, H_{\2}^{\Lambda} \wedge H_{\2}^{\Sigma}  - V \, \textrm{vol}_4  \nonumber  \ .
\end{eqnarray}
The scalar kinetic terms are given by a standard metric on (\ref{ScalManN=2}). The precise form of the minimal ($D\phi$, etc.) and non-minimal ($\mathcal{R}_{\Lambda\Sigma}$, $\mathcal{I}_{\Lambda\Sigma}$) couplings of the scalars to the vectors are not needed in this paper. We do need the expression of the scalar potential $V$, which fixes the radius $L$ of its AdS$_4$ vacua (for which $V_0 < 0$ at a critical point) as
\begin{equation} \label{eq:AdSLV}
L^2 = - \frac{6}{V_0} \; .
\end{equation}
The potential is different for each gauging. For the SU(3)--invariant sector of the purely electric SO(8) gauging \cite{deWit:1982ig}, the potential is \cite{Warner:1983vz}, in the conventions of \cite{Larios:2019kbw},
{\setlength\arraycolsep{1pt}
\begin{eqnarray}	\label{eq:scalarpotSU3inSO8}
g^{-2} V & = & -12 e^\varphi -6 e^{-2\phi -\varphi} XY \big( e^{4\phi} +Y^2+Z^2 \big) -12 e^\varphi (Y-1) \big( 1+Y -\tfrac32 XY \big)   \\[5pt]
&& +6 e^{-2\phi -\varphi} (Y-1) \big( e^{4\phi} +Y^2 +Z^2 \big) X^2  
 + e^{-3\varphi} \Big[ \tfrac12 e^{-4\phi} +a^2 -1 + \tfrac12 e^{4\phi} (1+a^2)^2 \nonumber \\
&& \qquad \qquad \quad  +\tfrac12 e^{-4\phi} (Y-1) \big( 1+2 Z^2 - 2 e^{4\phi} +Y(1+2 e^{4\phi} +2Z^2)+ Y^2 + Y^3  \big)  \Big] X^3 \nonumber \; .
\end{eqnarray}
}For the dyonic ISO(7) gauging the, SU(3)-invariant potential reads \cite{Guarino:2015qaa}
{\setlength\arraycolsep{1pt}
\begin{eqnarray}	\label{eq:scalarpotSU3inISO7} 
V &=& \frac{1}{2} g^2 \Big[X^3 e^{4 \phi -3 \varphi }+12 e^{2 \phi -\varphi }\Big(X^2 (Y-1)-X Y\Big) +12 e^{\varphi } \Big(3 X Y (Y-1)-2 Y^2\Big)\Big]	\nonumber\\[6pt]
	&&\, -g\, m\, \chi\,  e^{3 \varphi +2 \phi } \Big[6 (Y-1)+e^{2 \phi -2 \varphi }(X-1)\Big]+\frac{1}{2} m^2 e^{3 \varphi +4 \phi }\;.	
\end{eqnarray}
Finally, the SU(3)-invariant potential for the $[ \textrm{SO}(6) \times \textrm{SO}(1,1)] \ltimes\mathbb{R}^{12}$ gauging is
{\setlength\arraycolsep{1pt}
\begin{eqnarray}	\label{eq:scalarpotSU3inSO6}
V &=& 6\, g^2 e^{\varphi } \Big[3 X Y (Y-1)-2 Y^2\Big]+6\, g\, m\, \chi \, e^{3 \varphi +2 \phi } (Y-1) \Big[1-e^{-4 \phi } \left(Y^2+Z^2\right)\Big]	\nonumber\\[6pt]
	&&\,+\frac{1}{2} m^2 e^{3 \varphi } \Big[e^{4 \phi}+e^{-4 \phi } \left(Y^2+Z^2\right)^2-2 \left(Y^2-2 Y+Z^2\right)\Big]\; , 
\end{eqnarray}
}as follows from \cite{Guarino:2019oct}. In (\ref{eq:scalarpotSU3inSO8})--(\ref{eq:scalarpotSU3inSO6}), $g$ and $m$ are the electric and magnetic gauge couplings of the parent $\cN=8$ supergravities. For the latter two gaugings at hand, these can be set equal, $m=g$, without loss of generality \cite{Dall'Agata:2014ita}. This is in fact what we have done, following \cite{Inverso:2016eet,Guarino:2019oct}, to write the type IIB uplifts of sections \ref{sec:SpecIIB} and \ref{sec:IIBSO4Spectrum}.  We have also employed the shorthand notations \cite{Guarino:2015qaa}
\begin{equation} \label{scalDefs}
X \equiv 1+ e^{2\varphi} \chi^2
\hspace{5mm} , \hspace{5mm}
Y \equiv 1+\tfrac{1}{4} \, e^{2\phi} \, (\zeta^2+\tilde{\zeta}^2)
\hspace{5mm} , \hspace{5mm}
Z \equiv  e^{2\phi} \, a  \ .
\end{equation}

The AdS vacua of the SO(8), ISO(7) and $[ \textrm{SO}(6) \times \textrm{SO}(1,1)] \ltimes\mathbb{R}^{12}$ $\cN=8$ gaugings that preserve at least the SU(3) subgroup of those gauge groups were respectively investigated in \cite{Warner:1983vz,Guarino:2015qaa,Guarino:2019oct}. In our conventions, these correspond to extrema of the scalar potentials (\ref{eq:scalarpotSU3inSO8})--(\ref{eq:scalarpotSU3inSO6}). The location of these vacua in scalar space, in the notation that we are using, can be respectively found in table 2 of \cite{Larios:2019kbw}, table 3 of \cite{Guarino:2015qaa} (with labels $+$ there replaced with labels $v$ here), and table \ref{table:so6xso11criticalpoints} above.

For the SO(8) and ISO(7) gaugings, the G$_2$-invariant scalar sector (employed in the former context in section \ref{sec:G2sec} of the main text) is reached from the SU(3)-invariant sector through the identifications \cite{Larios:2019kbw,Guarino:2015qaa}
\begin{equation} \label{eq:G2sector}
\phi = \varphi \; , \qquad 
\tilde \zeta = -2\chi \; , \qquad 
a= \zeta = 0 \; .
\end{equation}
The SU$(4)_c$-invariant sector of the SO(8) gauging is retrieved via \cite{Larios:2019kbw}
\begin{equation} \label{SU4csector}
e^{-2\phi} = 1-\tfrac14 ( \zeta^2 + \tilde{\zeta}^2 ) \; ,  \quad a =0 \; ,  \quad 
e^{-2\varphi} = 1- \chi^2 \; .
\end{equation}

\bibliography{references}
\end{document}